\documentclass[12pt]{article}
\usepackage[margin=2cm]{geometry}
\usepackage[english]{babel}
\usepackage{graphicx,subfigure}
\usepackage{tamasty}
\begin{document}
%%%%%%%%%%%%%%%%%%%%%%%%%Command definitions%%%%%%%%%%%%%%%%%%%%%%%%%%%%
%%%%%%%%%%%%%%%%%%%%%%%%%%%%%%%%%%%%%%%%%%%%%%%%%%%%%%%%%%%%%%%%%%%%%%%%
%%%%%%%%%%%%%%%%%%%%%%%%%%Tama's%%%%%%%%%%%%%%%%%%%%%%%%%%%%%%%%%%%%%%%%
\newcommand{\ee}{\`e\ }
\renewcommand{\aa}{\`a\ }
\newcommand{\oo}{\`o\ }
\newcommand{\uu}{\`u\ }
\newcommand{\ket}[1]{\ensuremath {|\: #1 \: \rangle}}
\newcommand{\bra}[1]{\ensuremath{\langle \: #1 \:|}}
\newcommand{\braket}[2]{\ensuremath{\langle \: #1 \: | \: #2 \: \rangle}}
\newcommand{\ketbra}[2]{\ensuremath{| \: #1 \:\rangle \langle \: #2 \:  |}}
\newcommand{\ves}[2]{\ensuremath{#1_1,#1_2, \ldots, #1_{#2}}}
\newcommand{\ve}[2]{\ensuremath{#1(1),#1(2) \ldots, #1(#2)}}
\newcommand{\Hc}{\ensuremath{\mathcal{H}_{cursor\;}}}
\newcommand{\Hr}{\ensuremath{\mathcal{H}_{register\;}}}
\newcommand{\Hcr}{\ensuremath{\mathcal{K}_{cur\;}}}
\newcommand{\Hrr}{\ensuremath{\mathcal{K}_{reg\;}}}
\newcommand{\Hm}{\ensuremath{\mathcal{H}_{machine\;}}}
%%%%%%%%%%%%%%%%%%%%%%%%%%%%%%%%%%%%%%%%%%%%%%%%%%%%%%%%%%%%%%%%%%%%%%%%
%%%%%%%%%%%%%%%%%%%%%%%%%%%%%%%%%%%%%%%%%%%%%%%%%%%%%%%%%%%%%%%%%%%%%%%%
%%%%%%%%%%%%%%%%%%%%%%Cross references%%%%%%%%%%%%%%%%%%%%%%%%%%%%%%%%%%
\newcommand{\eref}[1]{(\ref{#1})}
\newcommand{\sref}[1]{section~\ref{#1}}
\newcommand{\fref}[1]{figure~\ref{#1}}
\newcommand{\tref}[1]{table~\ref{#1}}
\newcommand{\Eref}[1]{Equation (\ref{#1})}
\newcommand{\Sref}[1]{Section~\ref{#1}}
\newcommand{\Fref}[1]{Figure~\ref{#1}}
\newcommand{\Tref}[1]{Table~\ref{#1}}
\title {Speed and entropy of an interacting continuous time quantum walk}
\author{
\small {Diego de Falco}
\footnote{CIMAINA, Centro Interdisciplinare Materiali e Interfacce Nanostrutturati, Universit\aa degli Studi di Milano.}\\
\small{Dipartimento di Scienze dell'Informazione}\\
\small{Universit\aa degli Studi di Milano}\\
\small{via Comelico 33, 20135 Milano, Italy}.\\ 
\small{\emph{e-mail:} defalco@dsi.unimi.it}
\and
\small{Dario Tamascelli}\\
\small{Dipartimento di Matematica}\\
\small{Universit\aa degli Studi di Milano}\\
\small{via Saldini 50, 20133 Milano, Italy.}\\
\small{\emph{e-mail:} tama@mat.unimi.it}
}
\date{}
%\eads{\mailto{defalco@dsi.unimi.it}, \mailto{tama@mat.unimi.it}}
%
%
\maketitle

\begin{abstract}
We present some dynamic and entropic considerations about the evolution of a continuous time quantum walk implementing the clock of an autonomous machine. On a simple model, we study in quite explicit terms the Lindblad evolution of the clocked subsystem, relating the evolution of its entropy to the spreading of the wave packet of the clock. We explore possible ways of reducing the generation of entropy in the clocked subsystem, as it amounts to a deficit in the probability of finding the target state of the computation. We are thus lead to examine the benefits of abandoning some classical prejudice about how a clocking mechanism should operate.
\end{abstract}
%
%%%%%%%%%%FILES ARTICOLO%%%%%%%%%%%%%%
%%%%%%%%%%%%%%%%%%%%%%%%%%%%%%%%%%%%%%
\section{Introduction}
We study a model of quantum computation originally due to Feynman \cite{feyn86}, in which the evolution of the quantum register is controlled by an auxiliary clock or cursor register. Feynman pointed out the computational relevance, for such a system, of a Hamiltonian of the form
\begin{equation}\label{eq:hamiltoniana1}
	H = -\frac{\lambda}{2} \sum_{x,y \in G}\;U(x,y;\mathbf{\bsigma}) \tau_+(y) \tau_-(x)+ h.c.
\end{equation}
by showing that the CCNOT primitive (and therefore a universal quantum computer) can be implemented by a suitable choice of the graph $G$ and of the dependence on the \emph{register spins} $\bsigma$ of the dynamical variables $U$ which couple the \emph{cursor spin} creation and annihilation operators $\tau_{\pm}$.\\
A general architecture emerges from the above model in which an excitation of the $\btau$ field performs a quantum walk on a planar graph, the basic events being the flipping of a $\mathbf{\sigma}$ spin determined by the flowing of the cursor current (the NOT primitive) and the conditional choice, determined by the state of a $\mathbf{\sigma}$ qubit, of the flow of the cursor along alternative edges of the graph (the SWITCH primitive).\\
It is to the clock subsystem that we devote most of our attention in this paper, contributing an explicitly solvable toy model to the long-standing exploration \cite{salecker58} of the quantum limitations of time measurement (for a discussion of the foundational nature of this problem and for extensive updated references, see \cite{gambini04}). We study, furthermore, by frankly heuristic numerical means when necessary, variations on the theme of our toy model, with the purpose of gaining and sharing experience on a problem (optimizing the performance of a quantum clock) that might become of practical relevance in studying ballistic computation on nanostructures \cite{bertoni00,ionicioiu01}.\\
The dynamics under the Hamiltonian \eref{eq:hamiltoniana1} was examined by A. Peres \cite{peres85} on a linear chain with edges between nearest neighbor sites, with the  $U$'s  taken to be numerical functions of the positions; equivalently stated, the dynamics was studied along the linear chain of \emph{logical successors} of an initial \emph{register state}. Peres showed, in the suggestive terminology of the more recent Reference \cite{christandl04}, that unit \emph{fidelity} could be achieved in the transfer of a state along such a linear chain by suitable \emph{engineering} of the coupling constants $U$.\\
Interest in the case in which the  $U$'s are numerical functions (as opposed to matrices acting on additional spins) has been revived by the success of the quantum walk paradigm \cite{childs02}. In \cite{christandl04} the graph   is considered in its all important role as a wire able to spatially transfer a single-spin  quantum state; in \cite{albanese04} the interesting possibility is examined of performing the mirror inversion of a many-spin state.\\
For a recent presentation of the full model \eref{eq:hamiltoniana1}, which we call an \emph{interacting} $XY$ \emph{system}, we refer to \cite{defa04}, where particular attention is paid to the role of additional controlling spins in implementing successive visits to selected parts of the graph in iterated computations (quantum subroutines) and in the storage of results in telomeric chains.\\
In this note we examine an explicitly solvable, yet computationally non trivial, instance of the Hamiltonian \eref{eq:hamiltoniana1}, paying specific attention to non-positional observables of the system: our main concerns will be speed (of computation) and entropy (of controlled and/or controlling subsystem). We also consider the observable \emph{number of particles} (agents performing a quantum walk along the $XY$ chain), and discuss the interest and limitations of the proposal of a multi-hand quantum clock as a substitute for the loops implementing iterated applications of quantum subroutines.\\[5pt]
The paper is organized as follows. In Section 2 we present the model and establish our notation. In Section 3 we relate the speed of computation to the group velocity of the motion of the cursor wave packet along the graph. In Section 4 we discuss the build-up of entropy  in the \emph{clocked} subsystem because of  the spreading of the wave packet of the \emph{clocking agent}. An outline of possible choices of the initial form of this wave packet bringing such entropy build-up close to a minimum is given in Section 5. Section 6 examines the possible interest of \emph{multiagent} spin networks. \mbox{Section 7 is devoted to conclusions and outlook.}
\section{The model}
The model we consider consists of two functionally distinct parts, the \emph{input/output register}  and the \emph{clock}  or \emph{cursor}, and evolves as an \emph{autonomous system} under the sole effect of its initial state not being an eigenstate of the Hamiltonian of the composite system.\\
Let $\ket{R(1)}\in \Hr$ be the initial state of the input/output register.  Set
\begin{equation}
	d=dim(\Hr).
\end{equation}
Let \ves{U}{N-1} be the unitary operators representing the successive \emph{primitive} steps of the computation to be performed. Suppose, namely, that the goal of the computation is to transform the input state \ket{R(1)} into the output state $\ket{R(N)}= U_{N-1} \cdot \ldots U_2 \cdot U_1 \ket{R(1)}$   by visiting the successive intermediate states
\begin{equation}
	\ket{R(x)}= U_{x-1} \ket{R(x-1)},\ 1 < x < N.
\end{equation}
Following the approach of \cite{feyn86}, we model the clocking mechanism, which sequentially applies the transformations \ves{U}{N-1} to the register, with a quantum mechanical system, the \emph{cursor}.\\
We call \Hc the $s$-dimensional ($s \geq N$) state space of this system and refer it to a selected orthonormal basis $\ket{C(1)},\ \ket{C(2)}$,\ldots,\ \ket{C(s)}. It will help the intuition, and will explain the notation used below, to think of an explicit implementation of the cursor by $s$ spin $1/2$ particles and to think of the state \ket{C(x)} as obtained by flipping ``up'' the spin in position $x$ with respect to the ``all down'' reference state.\\
We suppose that the state of the overall system, the \emph{machine}, evolves in the Hilbert space $\Hm = \Hr \otimes \Hc$ under the action of a Hamiltonian of the form
\begin{equation}\label{eq:hamiltoniana2}
	H = -\frac{\lambda}{2} \sum_{x=1}^{s-1} U_x \otimes \ketbra{C(x+1)}{C(x)} + U_x^{-1} \otimes \ketbra{C(x)}{C(x+1)}.
\end{equation}
Notice that only \ves{U}{N-1} are assigned by the algorithm we are interested in; $U_N,\ldots, U_{s-1}$ are to be assigned as a part of the description of the clocking mechanism.\\
For instance, Reference \cite{apolloni02} presents the case in which $U_N,\ldots, U_{s-1}=I_r$, the identity in \Hr,  and shows the role of the cursor sites $N,\ldots,s$  as a storage mechanism of the output \ket{R(N)}.\\
An alternative point of view was taken in some of the numerical examples of \cite{defa04}, motivated by Grover's algorithm: one may suppose all of the $U_x$ to coincide, in such a context, with Grover's $estimation \cdot oracle$ step $G$, and study the effect of applying $G$ more than the optimal number $N-1$ of times.\\
This point of view will be taken also in some numerical examples of this paper, where we focus our attention on a Hamiltonian of the form:
\begin{equation} \label{eq:hamiltoniana3}
	H = -\frac{\lambda}{2} \sum_{x=1}^{s-1} U_x \otimes \tau_+(x+1) \tau_-(x) + U_x^{-1} \otimes \tau_+(x)\tau_-(x+1).
\end{equation}
Most of our numerical examples will refer in fact to the following particular instance (Toy model):
\begin{equation} \label{eq:hamiltoniana4}
		H_T = -\frac{\lambda}{2} \sum_{x=1}^{s-1} e^{-i\frac{\alpha}{2} \sigma_2} \tau_+(x+1) \tau_-(x) + e^{i\frac{\alpha}{2} \sigma_2}\tau_+(x)\tau_-(x+1).
\end{equation}
Here $\underline{\sigma}=(\sigma_1, \sigma_2, \sigma_3)$ is the single register spin $1/2$ that we are going to consider in our model; the cursor subsystem is implemented as a collection of spin $1/2$ systems $\underline{\tau}(j)=(\tau_1(j),\tau_2(j),\tau_3(j))$ , $j \in {1,\ldots,s}$, and $\tau_{\pm}(j)= (\tau_1(j)\pm i \tau_2(j))/2$.\\
The Hamiltonian \eref{eq:hamiltoniana2} can, of course, be considered as the restriction of \eref{eq:hamiltoniana3} to the eigenspace belonging to the eigenvalue $1$ of 
\begin{equation}
	N_3=\sum_{x=1}^s \frac{1+\tau_3(x)}{2},
\end{equation}
provided we identify $\ket{C(x)}$ with the simultaneous eigenstate of $\ve{\tau_3}{s}$ in which only  $\tau_3(x)$ has eigenvalue $+1$.

\section{Speed of computation} \label{sec:speed}
The eigenvalue problem for the Hamiltonian \eref{eq:hamiltoniana2} is solved by the following Ansatz for the eigenstates
\begin{equation}
	\ket{e}= \sum_{x=1}^{s-1} v(x)\  U_{x-1} \cdot \ldots \cdot U_1 \ket{r} \otimes \ket{C(x)},
\end{equation}
with \ket{r} any non vanishing vector in \Hr, suggested by the conservation laws discussed in \cite{peres85} (equation 19, p.3270). Inserting this Ansatz, the eigenvalue problem becomes:
\begin{equation}
	e \ v(x) = -\frac{\lambda}{2}(v(x+1)+v(x-1)). 
\end{equation}
It is immediate to recognize in the right hand side a finite difference approximation of the Laplace operator (the \emph{free} Schr\"odinger equation is, not surprisingly, at work in the motion of the clock); together with the boundary conditions $v(0)=v(s+1)=0$, this leads in an obvious way to the eigenvalues
\begin{equation}
	e_k = -\lambda \ \cos\left( \frac{k \pi}{s+1}\right),\ k=1,2,\ldots,s.
\end{equation}
The multiplicity of each eigenvalue is equal to $d=dim(\Hr)$. An orthonormal basis in the eigenspace belonging to the eigenvalue $e_k$ is given by:
\begin{equation} \label{eq:ekrj}
	\ket{e_k; r_j}= \sum_{x=1}^{s-1} v_k(x) U_{x-1}\cdot \ldots \cdot U_1 \ket{r_j} \otimes \ket{C(x)},
\end{equation}
where $\ket{r_1},\ldots,\ket{r_d}$ is an orthonormal basis in \Hr, and
\begin{equation} \label{eq:autofunzioni}
	v_k(x)= \sqrt{\frac{2}{s+1}} \sin\left( \frac{k \pi}{s+1} x \right).
\end{equation}
The same statements hold, of course, for the Hamiltonian \eref{eq:hamiltoniana3} in the eigenspace belonging to the eigenvalue $1$ of $N_3$.\\
An initial state (at time $t=0$) of the form
\begin{equation} \label{eq:condini1}
	\ket{M_1} = \ket{R(1)} \otimes \ket{C(1)}
\end{equation}
evolves, under the Hamiltonian \eref{eq:hamiltoniana2}, into
\begin{equation} \label{eq:evoluzione1}
	\ket{M_1(t)} = e^{-iHt} \ket{M_1}= \sum_{x=1}^s c(t,x;s)\ket{R(x)} \otimes \ket{C(x)}
\end{equation}
where
\begin{equation} \label{eq:c}
	c(t,x;s)= \frac{2}{s+1} \sum_{k=1}^s \exp\left[ i \lambda t \cos\left( \frac{k \pi}{s+1}\right)\right] \sin\left( \frac{k \pi}{s+1}\right) \sin\left( \frac{k \pi x}{s+1} \right).
\end{equation}
Equation \eref{eq:evoluzione1} singles out the interest of the observable
\begin{equation}
	Q = \sum_{x=1}^s x\;\ketbra{C(x)}{C(x)}= \sum_{x=1}^s x \frac{1+\tau_3(x)}{2},
\end{equation}
the \emph{position of the cursor}: it shows that \underline{if},  at any time $t$, $Q$ is measured on a system in the state \ket{M_1(t)} \underline{and} the the result $x$ is observed, \underline{then} the register collapses into the state \ket{R(x)} obtained from \ket{R(1)} by the application of \ves{U}{x-1} in the right order.\\
The observable $Q/t$ acquires thus the meaning of number of primitives per unit time applied to the initial condition \ket{R(1)} in the time interval $(0,t)$. In order to study the behavior over long intervals of time ($t \rightarrow +\infty$) of this observable in the case of a long computation ($s \rightarrow +\infty$) it is expedient to study its characteristic function
\begin{equation} \label{eq:fcaratteristica}
	\phi_{s,t}(z) = \bra{M_1(t)} \exp \left(iz \frac{Q}{t}\right) \ket{M_1(t)} = \sum_{x=1}^s |c(t,x;s)|^2 \exp \left(iz \frac{x}{t} \right),
\end{equation}
namely the Fourier transform of its probability distribution.\\
The large $s$ behavior is easily studied by inserting the explicit integral representation of the $s \rightarrow +\infty$ limit of  \eref{eq:c} into \eref{eq:fcaratteristica} ; the $t \rightarrow +\infty$ limit is similarly studied by substituting the sum over $v=x/t$, step $1/t$,  appearing in \eref{eq:fcaratteristica} with an integral and evaluating the leading contributions by a standard stationary phase argument.  We thus obtain:
\begin{equation} \label{eq:limitecaratteristica}
	\lim_{t \rightarrow \infty} \lim_{s \rightarrow \infty} \bra{M_1(t)} \exp \left( iz\frac{Q}{t} \right) \ket{M_1(t)} = \int_0^1 \frac{4 v^2}{\pi \sqrt{1-v^2}} e^{izv} dv.
\end{equation}
As convergence in the sense of characteric functions implies convergence in the sense of cumulative distribution functions (\emph{convergence in law}),  we conclude that a ``long''  computation starting from the initial condition \eref{eq:condini1} proceeds ``in the long run'' at a rate of $V(M_1)$ steps per unit time (the unit of time having been set so that $\lambda=1$ ),  $V(M_1)$  being the random variable defined by having as its characteristic function the right hand side of  \eref{eq:limitecaratteristica}; equivalently stated it has probability density 
\begin{equation} \label{eq:densitav}
	f_{V(M_1)}(v)= I_{(0,1)}(v)  \frac{4 v^2}{\pi \sqrt{1-v^2}}
\end{equation}
Here and in what follows we denote by $I_{(a,b)}$the indicator function of an interval $(a,b)$:
\begin{equation}
	I_{(a,b)}(x)= \left \{ 
	\begin{array}{c r}
	1 & \mbox{if $x \in (a,b)$}\\
	0 & \mbox{otherwise.} 
	\end{array}\right.
\end{equation}
The mean value
\begin{equation} \label{eq:EM1}
	E(V(M_1))= \int_0^1 v\; f_{V(M_1)}(v) dv = \frac{8}{3 \pi}
\end{equation}
and the variance
\begin{equation} \label{eq:varM1}
	var(V(M_1))=E\left((V(M_1))^2\right)-E\left(V(M_1)\right)^2=\frac{3}{4}- \left(\frac{8}{3 \pi} \right)^2
\end{equation}
are then easy to compute from \eref{eq:densitav}.\\
More generally,  for any positive integer $x_0$, a state such as 
\begin{equation} \label{eq:condini2}
	\ket{M_{x_0}}=\ket{R(x_0)} \otimes \ket{C(x_0)}
\end{equation}
having at a certain instant the cursor in $x_0$, evolves with a speed $V(M_{x_0})$ having cumulative distribution function
\begin{eqnarray}
F_{V(M_{x_0})}(v) & \equiv & Prob(V(M_{x_0}) \leq v) =  \\
& = & I_{(0,1)}(v)\left( \frac{2 \arcsin(v)}{\pi} - \frac{\sin(2 x_0 \arcsin(v))}{\pi x_0}\right)+I_{(1,+\infty)}(x) \nonumber
\end{eqnarray}
and expectation value
\begin{equation} \label{eq:EMx0}
	E(V(M_{x_0}))=\frac{8}{4\pi-\pi/x_0^2}
\end{equation}
Comparison between \eref{eq:EM1} and \eref{eq:EMx0} shows the effect of a measurement of $Q$. If, at a given $t$, $Q$ is measured and the result $x_0$ is found, then the state \eref{eq:evoluzione1}, into which the initial condition \eref{eq:condini1} has evolved, collapses into the state \eref{eq:condini2}. From this moment on the computation proceeds at the mean rate \eref{eq:EMx0}: for large values of $t$, reading the \emph{clock} is likely to reduce the speed of further computation by a factor $3/4$ (without, because of \eref{eq:ekrj}, altering its correctness).  
\section{Entropy}
Motivated by the experience gained under the particular initial conditions \eref{eq:condini1} and \eref{eq:condini2} we define, for any (unentangled) initial condition of the form (for fixed $\epsilon \geq 1$)
\begin{equation} \label{eq:condini3}
	\ket{R; \psi_0}= \ket{R} \otimes\sum_{x=1}^{\epsilon} \psi_0(x) \ket{C(x)},
\end{equation}
the ``time-of-flight speed'' \cite{feyn65} of computation in the state $\psi_0$ as the random variable $V(\psi_0)$ having characteristic function
\begin{equation} \label{eq:fcaratteristica2}
	\phi_{V(\psi_0)}(z) = \lim_{t \to +\infty} \lim_{s \to +\infty} \bra{R; \psi_0} e^{itH}\exp \left(iz \frac{Q}{t}\right) e^{-itH}\ket{R; \psi_0}.
\end{equation}
The above limit is easily shown to exist by the techniques outlined in the previous section; it corresponds to the probability density
\begin{equation} \label{eq:densitapsi0}
	f_{V(\psi_0)}(v)=I_{(0,1)}(v) \frac{|\Psi(\arcsin(v))|^2+|\Psi(\pi-\arcsin(v))|^2}{\sqrt{1-v^2}}
\end{equation}
where
\begin{equation}
	\Psi(p)=\sqrt{\frac{2}{\pi}} \sum_{x=1}^{\epsilon} \sin(px) \psi_0(x).
\end{equation}
The observable $Q$ retains in this context the meaning of relational time \cite{gambini04} in the sense that,  \underline{given} that at any \emph{parameter time} $t$ the cursor is found at $x$, it is \underline{then} certain that the register is found in the state $U_{x-1}\cdot \ldots \cdot U_2 \cdot U_1 \ket{R}$.\\
In reading the output at any time $t$, namely in  the measurement of any, however carefully chosen,  observable of the register, there is an intrinsic uncertainty corresponding to the uncertainty about how far the computation has proceeded. The fact that $Q/t$ has a non trivial limit in law means that the leading term of the variance of $Q$ is proportional to $t^2$ and therefore that the uncertainty increases with $t$. This section is devoted to the examination of an example in which the notion of ``the most careful choice'' of the observable to read on the register can be made precise and shown to be pertinent to the algorithm considered.\\
We consider for the moment the initial condition \ket{M_1} given in \eref{eq:condini1} and its  time evolution \ket{M_1(t)} described in \eref{eq:evoluzione1}. More general initial conditions of the form  \eref{eq:condini3} will be examined in the next section.\\
Call
\begin{equation}
	\rho_m(t)=\ketbra{M_1(t)}{M_1(t)}
\end{equation}
the density matrix of the machine at time $t$.\\
By taking the partial trace $Tr_{\Hc}(\rho_m(t))$ with respect to the cursor degrees of freedom, we get the density matrix $\rho_r(t)$ of the register:
\begin{equation} \label{eq:matricereg}
	\rho_r(t)=\sum_{x=1}^s |c(t,x;s)|^2 \ketbra{R(x)}{R(x)}.
\end{equation}
Call $\lambda_j(t)$ the positive eigenvalues of $\rho_r(t)$  and \ket{b_j(t)} the corresponding eigenstates. A simple computation, amounting to the Schmidt decomposition \cite{peres93} of the state \eref{eq:evoluzione1}, shows, then, that the density matrix of the cursor is given by
\begin{equation} \label{eq:matricecur}
	\rho_c(t)= \sum_j \lambda_j(t) \ketbra{d_j(t)}{d_j(t)}
\end{equation}
where
\begin{equation}
	\ket{d_j(t)}= \frac{1}{\sqrt{\lambda_j(t)}} \sum_{x=1}^s c(t,x;s) \braket{b_j(t)}{R(x)}\> \ket{C(x)}.
\end{equation}
Because of \eref{eq:matricecur} and of the orthonormality of the states \ket{d_j(t)}, the von Neumann entropy of the register and also of the cursor is then given by
\begin{equation}
	S(\rho_c(t))=- \sum_j \lambda_j(t) \ln \lambda_j(t)= S(\rho_r(t)).
\end{equation}
We observe that, as \eref{eq:matricereg} shows, the von Neumann entropy of each subsystem does depend on the algorithm being performed. It is, indeed, \underline{only} under the hypothesis, nowhere made above, that the states  \ket{R(x)} are orthonormal that \eref{eq:matricereg} is the spectral decomposition of $\rho_r(t)$  (the von Neumann entropy becoming in this case equal to the Shannon entropy of the distribution of $Q$).\\
We focus our attention, in what follows, on our Toy model \eref{eq:hamiltoniana4}, in which the register is a single spin $1/2$ system. We indicate by $\underline{e_1},\underline{e_2},\underline{e_3}$ the versors of the three coordinate axes to which the components $\underline{\sigma}=(\sigma_1,\sigma_2,\sigma_3)$ of such a spin are referred.\\
In the basis \ket{\sigma_3=\pm 1}, the density operator $\rho_r(t)$ will be represented by the matrix
\begin{equation} \label{eq:rhor}
\rho_r(t)=\frac{1}{2} \left (
\begin{array}[pos]{c c}
1+s_3(t) & s_1(t)-i\>s_2(t)\\
s_1(t) + i\>s_2(t)  & 1-s_3(t)
\end{array}
\right )
\end{equation}
where
\begin{equation}
s_j(t)=Tr \left ( \rho_r(t) \cdot \sigma_j \right ),\;j=1,2,3.
\end{equation}
Equivalently stated, the \emph{Bloch representative} of the state $\rho_r(t)$ is given by the three-dimensional real vector
\begin{equation}
\underline{s}(t) = \sum_{x=1}^s  \left | c(t,x;s)  \right | ^2 \bra{R(x)} \underline{\sigma} \ket{R(x)}.
\end{equation}
We shall assume, in what follows, that the initial state of the cursor is \ket{C(1)} and that the initial state of the register is of the form
\begin{equation} \label{eq:condinigrover}
	\ket{R(1)}=\cos \left ( \frac{\theta}{2} \right ) \ket{\sigma_3=+1} + \sin \left(  \frac{\theta}{2} \right ) \ket{\sigma_3=-1}
\end{equation}
namely the eigenstate belonging to the eigenvalue $+1$ of $\underline{n}(1) \cdot \underline{\sigma}$, with
\begin{equation}
\underline{n}(1)= \underline{e}_1 \sin\theta + \underline{e}_3 \cos \theta. 
\end{equation}
We wish to remark that the above example captures the geometric aspects not only of such simple computational tasks as $NOT$ or $\sqrt{NOT}$ (viewed as rotations of an angle $\pi$ or $\pi/2$ respectively, decomposed into smaller steps of amplitude $\alpha$) but also of Grover's quantum search \cite{grover96}. If, indeed, the positive integer $\mu$ is the length of the marked binary word to be retrieved, and we set
\begin{equation}
\chi(\mu)=\arcsin(2^{-\frac{\mu}{2}})
\end{equation}
and
\begin{equation} \label{eq:theta}
\theta= \pi-2\>\chi(\mu)
\end{equation}
then the state \eref{eq:condinigrover} correctly describes the initial state $\ket{\iota}$ of the quantum search as having a component $2^{- \mu/2}$ in the direction of the target state, here indicated by $\ket{\omega}=\ket{\sigma_3=+1}$, and a component $\sqrt{1-2^{-\mu}}$ in the direction of the flat superposition, here indicated by $\ket{\sigma_3=-1}$, of the $2^\mu-1$ basis vectors orthogonal to the target state.
In this notations, if
\begin{equation} \label{eq:alpha}
\alpha=-4\>\chi(\mu),
\end{equation}
then the unitary transformation $\exp(-i \> \alpha\> \sigma_2/2)$ corresponds to the product $B\cdot A$ of the \emph{oracle} step
\begin{equation} \label{eq:oracle}
A=I_r-2 \> \ketbra{\omega}{\omega}
\end{equation}
and the \emph{estimation} step
\begin{equation}\label{eq:estimation}
B=2\>\ketbra{\iota}{\iota}-I_r.
\end{equation}
We refer the reader to the beautifully pedagogical approach of Jozsa \cite{jozsa99} where it is shown that in Grover's search the $\mu$-qubits register evolves in the \emph{two dimensional} space spanned by the its initial state \ket{\iota} and the target state \ket{\omega}. Thus, one qubit suffices to represent all instances of quantum search.\\
It is having in mind the connection with Grover's algorithm that, for the sake of definiteness, in the examples that follow we are going to consider the one-parameter family of models, parametrized by the positive integers $\mu$, corresponding to the choice \eref{eq:theta} and \eref{eq:alpha} of the parameters $\theta$ and $\alpha$ and to the choice $s=2^\mu+1$ of the number of cursor sites, corresponding to the possibility of performing up to an exhaustive search.\\
In the example defined by the above conditions it is
\begin{equation}
\bra{R(x)}\> \underline{\sigma} \ket{R(x)} = \sin \left ( \theta+(x-1)\alpha \right)\> \underline{e}_1+\cos \left ( \theta+(x-1)\alpha \right)\> \underline{e}_3
\end{equation}
and, therefore,
\begin{equation} 
\underline{s}(t) = \sum_{x=1}^s \left |c(t,x;s) \right | ^2  \left( \sin \left ( \theta+(x-1)\alpha \right)\> \underline{e}_1+\cos \left ( \theta+(x-1)\alpha \right)\> \underline{e}_3 \right ).
\end{equation}
Figure 1 presents, inscribed in the unit circle, a parametric plot of $(s_1(t),s_3(t))$ under the above assumptions .
\begin{figure}[t] 
	\centering
		\includegraphics{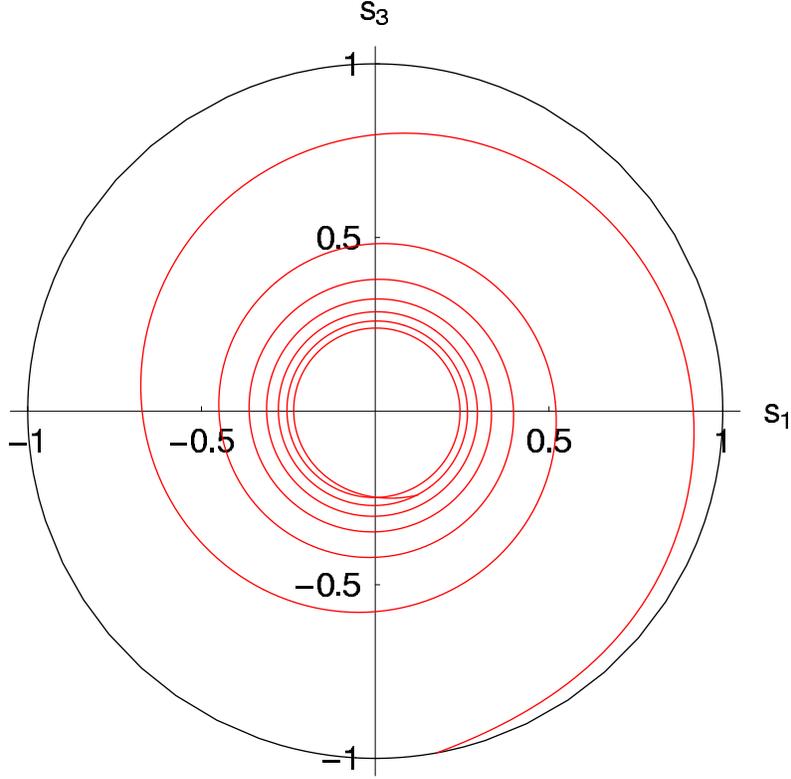}
\caption{A parametric plot of $\left (s_1(t),s_3(t)\right )$ for \mbox{$0 \leq t < s$}, $\lambda=1$. The choice $\mu=7,\;\chi= \arcsin(1/2^{\mu/2}),\; s=2^\mu +1,\; \alpha = -4 \chi,$ \mbox{$ \theta= \pi -2 \chi$} of the parameters is motivated by the connection with Grover's algorithm. Only the initial state lies on the unit circumference, the locus of pure states.}
\label{fig:figura1}
\end{figure}
It is convenient to describe the Bloch vector $\underline{s}(t)= s_1(t)\> \underline{e}_1+ s_3(t)\> \underline{e}_3$ in polar coordinates as
\begin{equation} \label{eq:toybloch}
\begin{array}{lcr}
s_1(t)=r(t)\sin{\gamma(t)},
& &
s_3(t)=r(t)\cos{\gamma(t)}.
\end{array}
\end{equation}
A very simple approximate representation of $\underline{s}(t)$ becomes then possible:
\begin{eqnarray}
r(t)e^{i \gamma(t)} & = & \sum_{x=1}^s |c(t,x;s)|^2 \exp(i(\theta+(x-1)\alpha)) = \nonumber \\
& = & \exp(i (\theta - \alpha))\sum_{x=1}^s |c(t,x;s)|^2 \exp(i\alpha x) \approx \nonumber \\
& \approx &  \exp(i (\theta - \alpha)) E\left( \exp(i \alpha \lambda t V(M_1(0))) \right)
\end{eqnarray}
The last step, legitimate for $1 << \lambda\; t <s$, requires only the explicit computation of the characteristic function corresponding to the probability density \eref{eq:densitav}, which leads to
\begin{equation}
	r(t)e^{i \gamma(t)} \approx \frac{2 \exp(i(\theta-\alpha))}{T}\left((J_1(T)- T \, J_2(T)+i (T \,H_0(T)-H_1(T))) \right)
\end{equation}
where $J_k$ and $H_k$ are, respectively, Bessel functions and Struve functions \cite{bessel}, and $T=\alpha \lambda t$.\\
The time evolution of the register subsystem is summarized by the Lindblad equation \cite{gorini76,lindblad76}
\begin{equation} \label{eq:lindblad}
	\frac{d \rho_r(t)}{dt} = -\frac{i}{2} \frac{d \gamma(t)}{dt} \left[ \sigma_2, \rho_r(t) \right]+ \frac{1}{4} \frac{d \ln r(t)}{dt} \left[\sigma_2 ,\left[ \sigma_2, \rho_r(t) \right] \right].
\end{equation}
The commutator term $\left[ \sigma_2, \rho_r(t) \right]$ describes the Hamiltonian part of the dynamics (after all we are considering a rotation about the $x_2$ axis); the double commutator $\left[\sigma_2 ,\left[ \sigma_2, \rho_r(t) \right] \right]$ describes, in much the same sense as equation 2.8 of \cite{milburn91}, the decohering effect of this rotation being administered by the cursor in discrete steps at random times.\\
The eigenvalues of $\rho_r(t)$ can be written as
\begin{equation} \label{eq:eigenvalues1}
\begin{array}{lcr}
\lambda_1(t)=\frac{1}{2} (1+r(t)), 
& &
\lambda_2(t)=\frac{1}{2} (1-r(t)).
\end{array}
\end{equation}
The von Neumann entropy $S\left( \rho_r(t) \right)$ is therefore
\begin{equation} \label{eq:vne}
S\left( \rho_r(t) \right) = - \frac{1+r(t)}{2} \ln \frac{1+r(t)}{2} - \frac{1-r(t)}{2} \ln \frac{1-r(t)}{2}.
\end{equation}
An example of its behaviour is shown in \fref{fig:figura2}.
\begin{figure}[t] 
	\centering
		\includegraphics{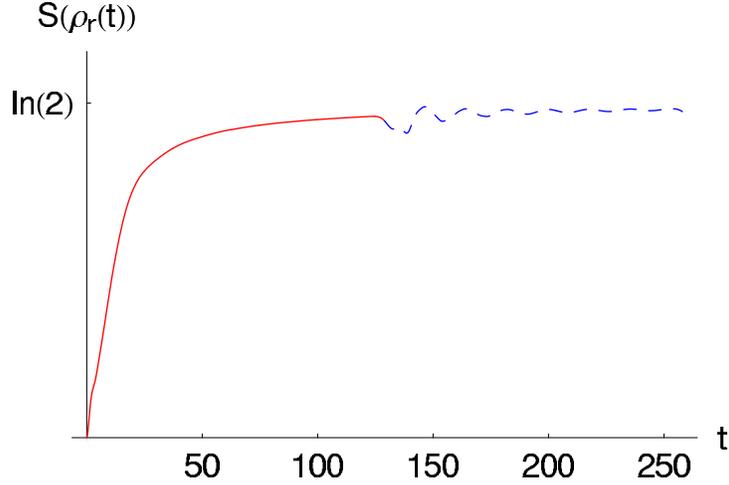}
		\caption{The von Neumann entropy of the register as a function of time, 
for the same model as in \fref{fig:figura1},  for $ 0 \leq t <s$  (solid line) and for  $ s \leq t < 2s$  (dashed line).}
\label{fig:figura2}
\end{figure}
The  eigenvectors corresponding to the eigenvalues \eref{eq:eigenvalues1} are, respectively
\begin{equation}
	\begin{array}{lcr}
		\ket{b_1(t)}= 
		\left (
				\begin{array}{c}
				\cos(\gamma(t)/2)\\
				\sin(\gamma(t)/2)
				\end{array}
		\right ),
		& &
		\ket{b_2(t)}= 
		\left (
				\begin{array}{c}
				-\sin(\gamma(t)/2)\\
				\cos(\gamma(t)/2)
				\end{array}
		\right ).
	\end{array}
\end{equation}
It is to be stressed that, at each time $t$, the projector \ketbra{b_1(t)}{b_1(t)} is, among the projectors on the state space of the register, the one having in the state  $\rho_r(t)$ the greatest probability of assuming, under measurement, the value $1$. Thus, the most careful choice (the one affected by minimum uncertainty) of the observable to read on the register at time $t$ is the projector \ketbra{b_1(t)}{b_1(t)}. In the case of Grover's algorithm one must measure the projector $\ketbra{\omega}{\omega}=\ketbra{\sigma_3=+1}{\sigma_3=+1}$ (and one easily can, because of the kickback mechanism analyzed, for instance, in \cite{defa04}) and has the freedom of choosing the time $\tau$ at which to perform the measurement. The best choice is therefore such that $\ket{b_1(\tau)}=\ket{\omega}$  (in our notational setting, $\tau$ is the time at which the helix of \fref{fig:figura1} crosses for the first time the positive $s_3$ axis). In spite of the fact of being now in the most favorable setting, one has, nevertheless, a deficit  $1-\lambda_1(\tau)$ in the probability of finding the target state.\\
As \fref{fig:figura3} shows,
\begin{figure}[!ht]
	\centering
		\includegraphics{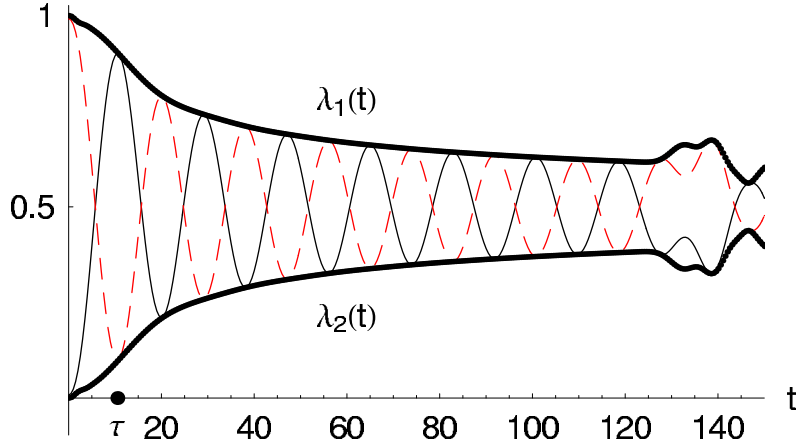}
		\caption{The same model as in \fref{fig:figura1} and \fref{fig:figura2}; $0 \leq t <1.2\,s$. The thin solid line is a graph of $Tr(\rho_r(t)\cdot (I_r+\sigma_3)/2)$, the probability of observing the target state $\ket{\omega}=\ket{\sigma_3=+1}$ in the example of Grover's algorithm. The dashed line is a graph of $Tr(\rho_r(t)\cdot (I_r-\sigma_3)/2)$, the probability of observing the ``undesired'' output $\ket{\sigma_3=-1}$. The upper and lower bounds on the probability of observing the target state are represented by the thick solid lines $\lambda_1(t)$ and $\lambda_2(t)$.}
	\label{fig:figura3}
\end{figure}
there are successive instants of time at which the probability of successful retrieval has a local maximum (a remnant of the periodic nature of Grover's algorithm when applied by an outside macroscopic agent) but the heights of these successive maxima form a sequence having a decreasing trend.\\
Further insight into our toy model is gained by examining the $t$ dependence of \mbox{$E(Q(t)) = \bra{M_1(t)}Q \ket{M_1(t)}$} and of the angle of polarization $\gamma(t)$. The example of \fref{fig:figuraeq}.a suggests that the mean value of speed derived from asymptotic considerations correctly describes the average behavior of the ``clocking'' subsystem also for finite values of  $0<t<s$. As \fref{fig:figuraeq}.b shows, the ``clocked'' subsystem system $\bsigma$ is, in turn, driven, on the average, into uniform rotational motion.
\begin{figure}[htbp]
	\centering
	\subfigure[]{\includegraphics[width=7cm]{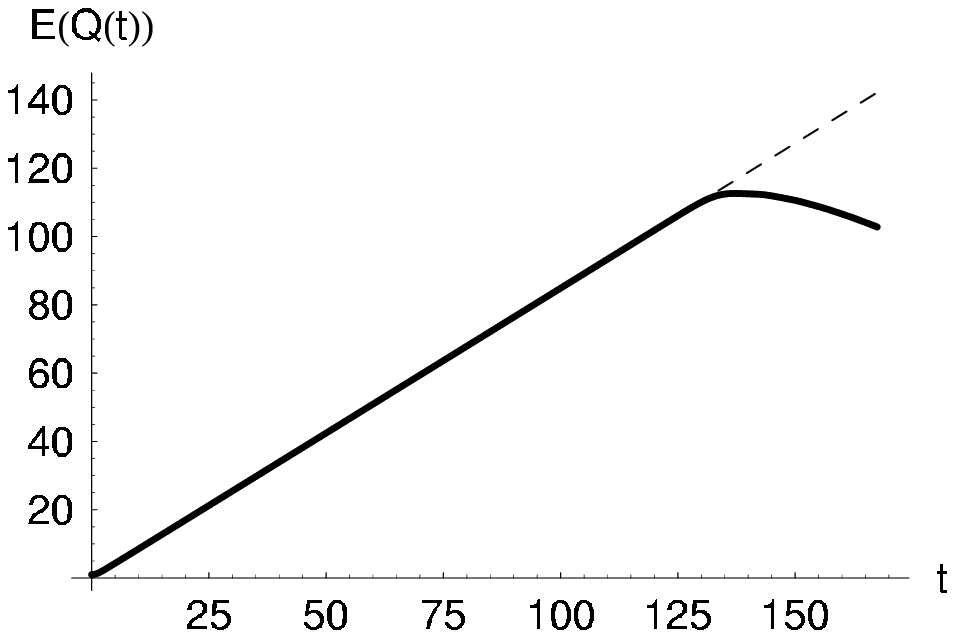}}
	\hspace{1cm}
	\subfigure[]{\includegraphics[width=7cm]{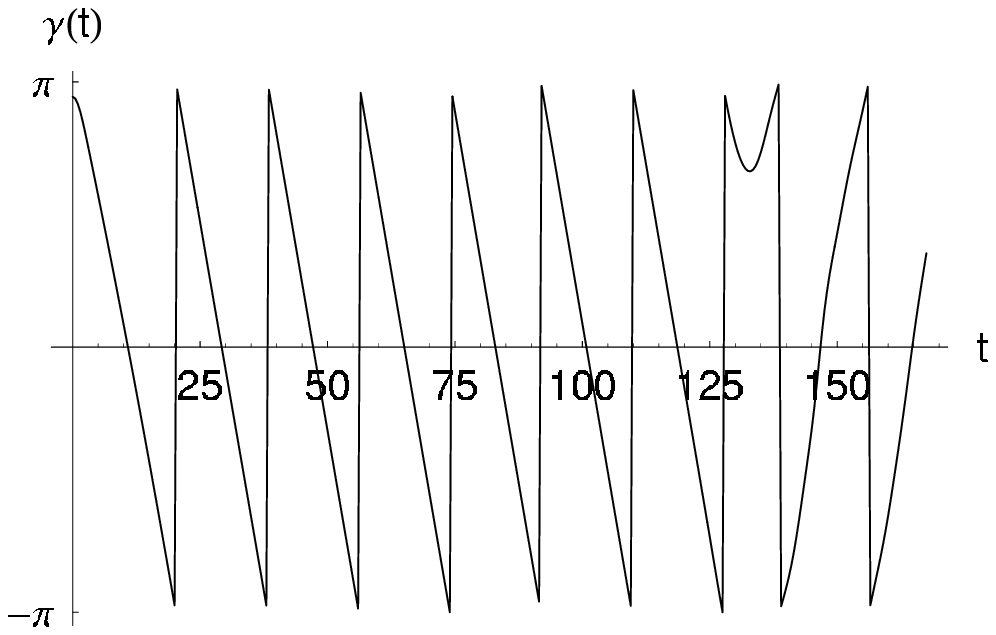}}
	\caption{Same parameters as in \fref{fig:figura1}. (a) $E(Q(t))$ (solid line) compared with the dashed straight line of slope $8/(3\pi)$. (b) The polar angle  $\gamma(t)$ of the Bloch vector \eref{eq:toybloch} as a function of $t$.}
\label{fig:figuraeq}
\end{figure}
\\Our model is so simple that we can explicitly study how the above semiclassical picture (in which the time parameter $t$ acquires operational meaning from its linear relation with mean values of configurational observables of clocking and clocked subsystem) is distorted by a measurement performed on either subsystem. The observations made at the end of the previous section about the effect of \emph{reading the clock} can indeed be complemented by the examination of the effect of \emph{reading the register}.\\
Suppose that the observable $\sigma_3$ has been measured at time $\tau$ and the result $+1$ has been found: the Bloch diagram of \fref{fig:figura20}.a shows then that the evolution of the register proceeds in much the same way as in the undisturbed situation of \fref{fig:figura1} (with the only obvious difference that the post-measurement initial condition  \ket{b_1(\tau)} lies  on the unit circumference).
\begin{figure}[ht]
\subfigure[]{\includegraphics[width=59mm]{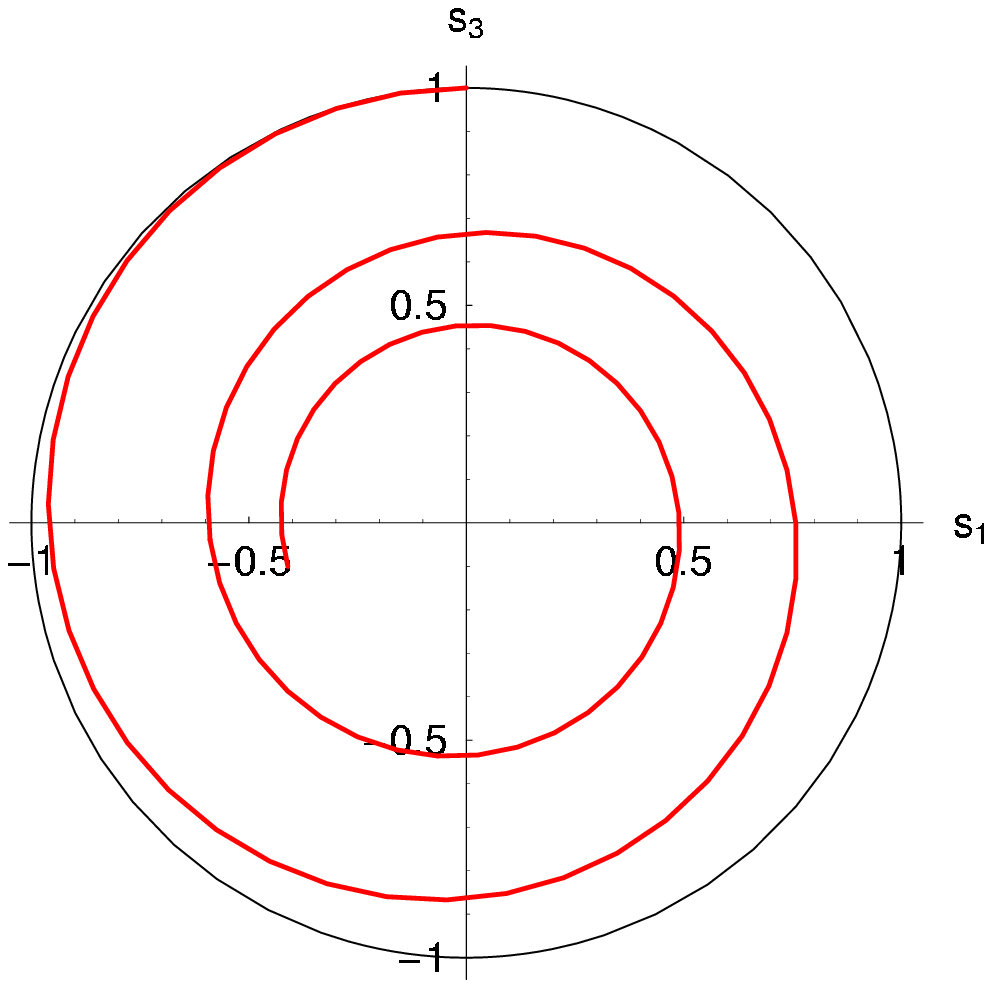}}
\hspace{2cm}
\subfigure[]{\includegraphics[width=59mm]{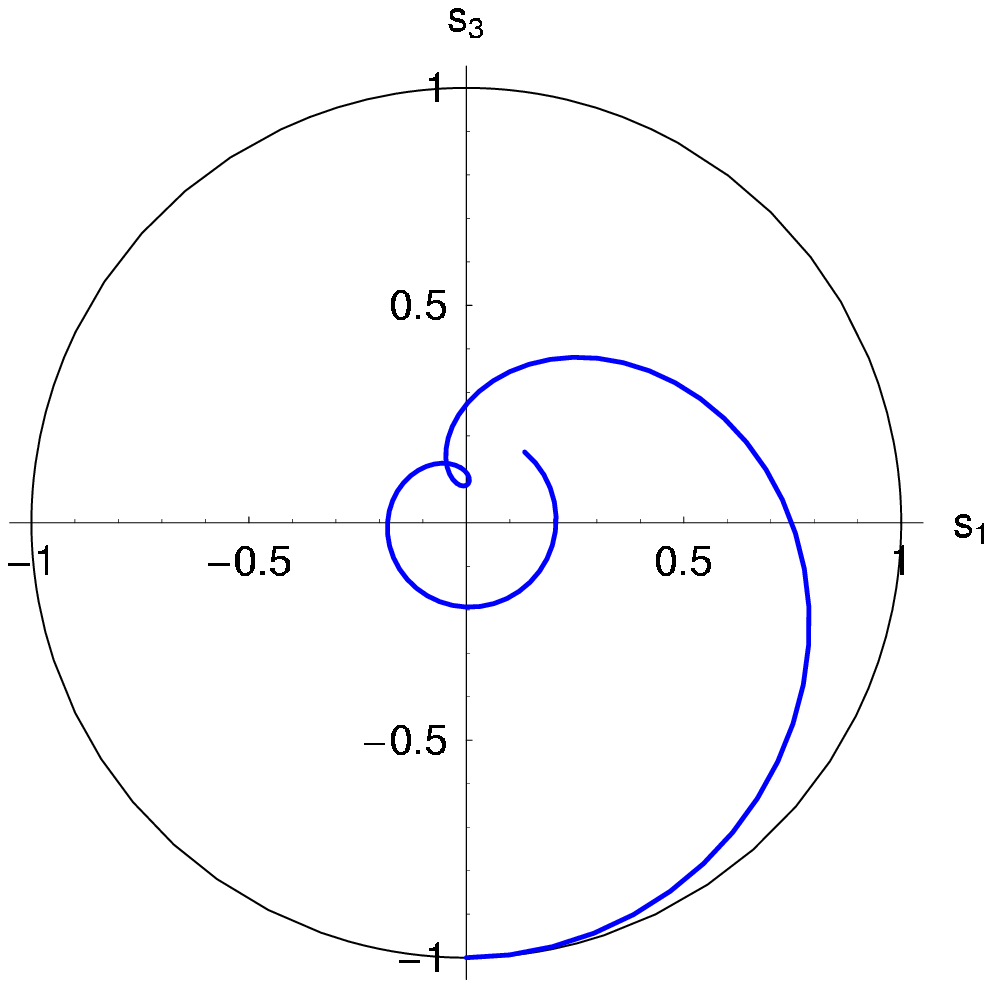}} \\
\centering
\subfigure[]{\includegraphics[width=79mm]{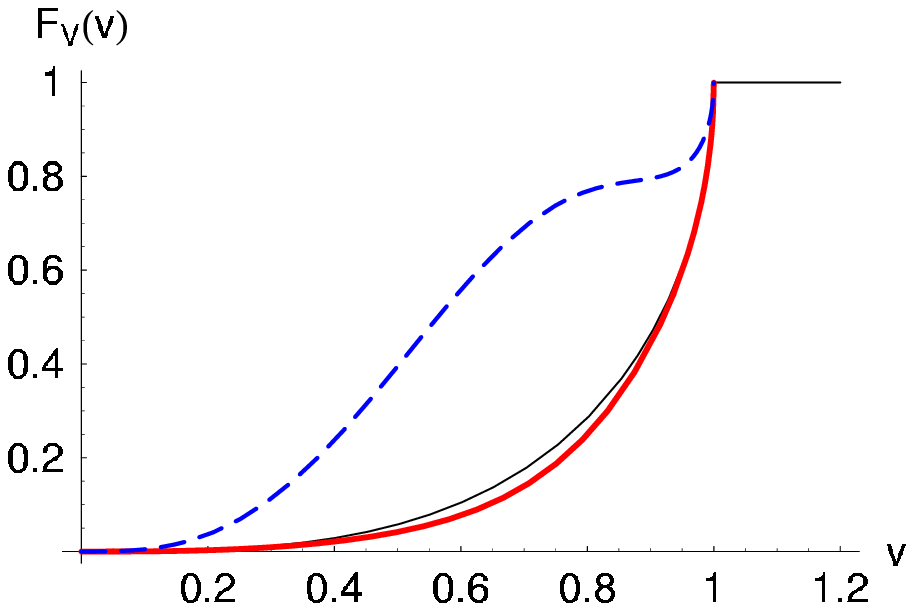}} 
\caption{Same parameters as in \fref{fig:figura1}; measurement of the observable $\sigma_3$  at time $\tau$. Frames (a) and (b) represent the evolution for $\tau < t \leq 4\tau$ of the Bloch vector when measurement returns $+1$ and $-1$ respectively. Frame (c) represents the cumulative distribution functions of the speed $V$ when the results $+1$ (solid thick line) and $-1$ (dashed line) have respectively been found; the solid thin line represents the c.d.f. of $V$ in case of no measurement.}
\label{fig:figura20}
\end{figure}
If, instead, the result $-1$ has been found (\fref{fig:figura20}.b) the post-measurement evolution of the register is completely different from the unperturbed one.\\
We conclude this section with an example of the insight that the time evolution of $S(\rho_r(t))$ can give on the algorithm \ves{U}{s-1} being performed by the machine. Suppose of using, instead of the assignment \eref{eq:hamiltoniana3} of the primitive steps, \mbox{$U_1=U_2=\ldots=U_{s-1}=\exp(-i \alpha \sigma_2/2)$}, the alternative assignment
\begin{equation} \label{eq:Ux}
	U_x= \left \{ 
	\begin{array}{c r}
	A & \mbox{for odd $x$}\\
	B & \mbox{for even $x$} 
	\end{array}\right.
\end{equation}
\begin{figure}[ht]
\subfigure[]{\includegraphics[width=79mm]{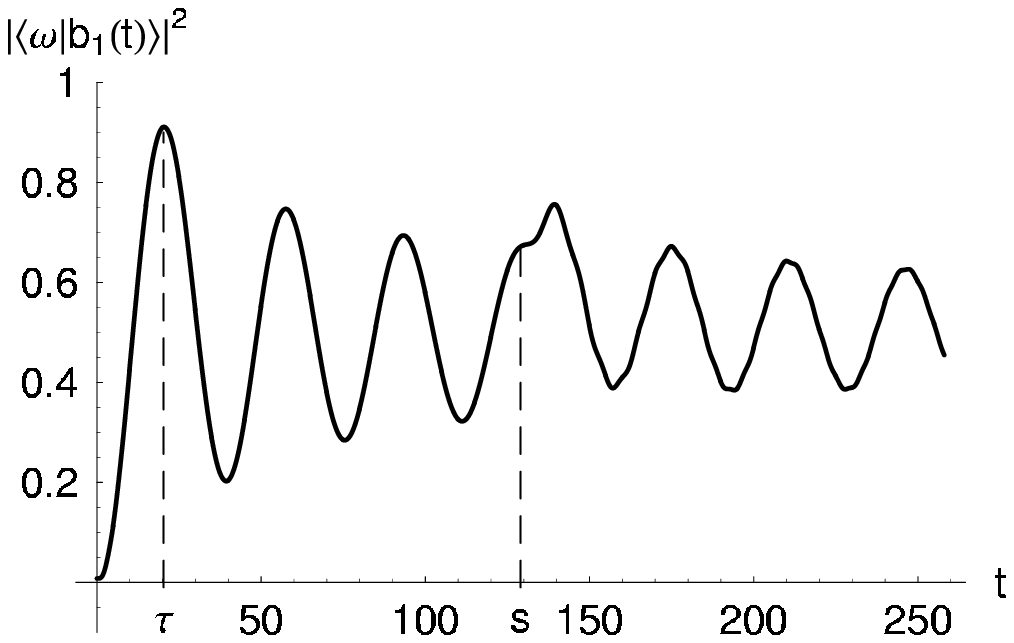}}
\subfigure[]{\includegraphics[width=79mm]{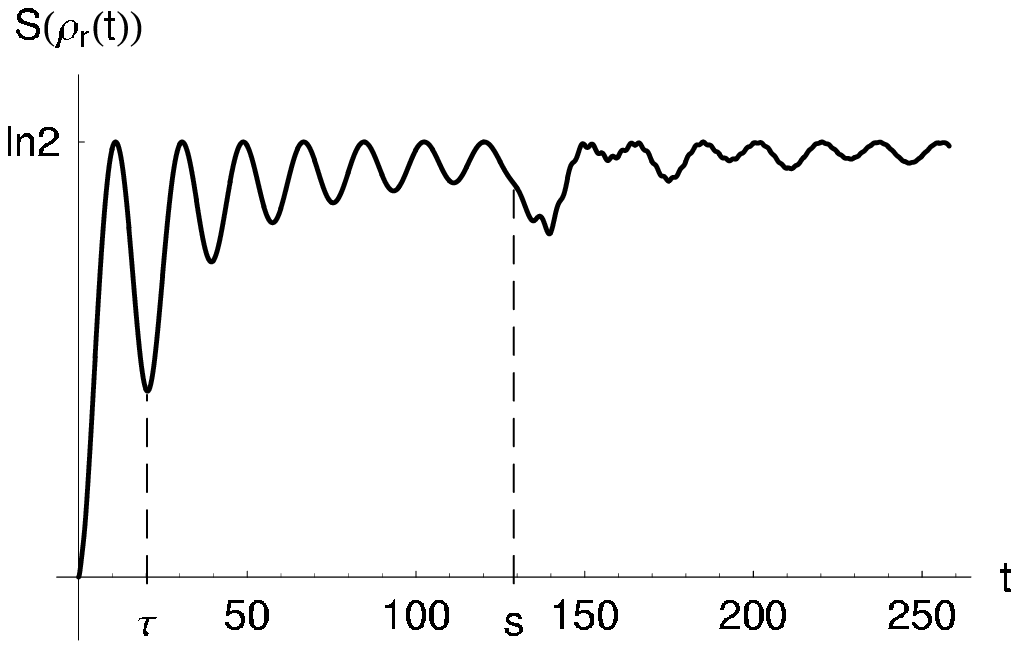}} \\
\centering
\subfigure[]{\includegraphics[width=59mm]{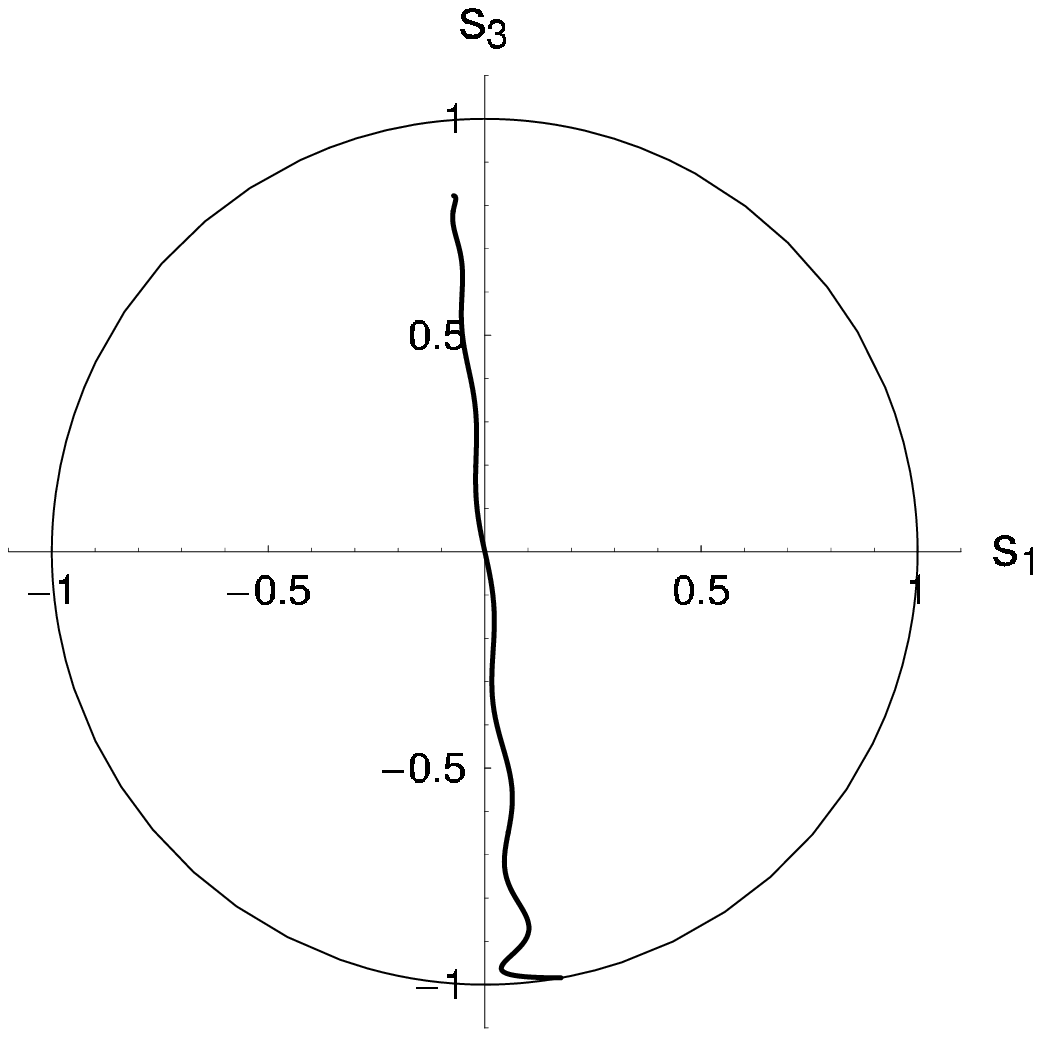}} 
\caption{Same parameters as in the previous figures; $U_x$ as in \eref{eq:Ux}. The Bloch diagram refers to the time interval $(0,\tau)$ needed to reach the first maximum in probability.}
\label{fig:figura4}
\end{figure}
where $A$ and $B$ are given by \eref{eq:oracle} and \eref{eq:estimation}. \Fref{fig:figura4} gives, for this example, a full account of the diffusive character \cite{grover01} of Grover's quantum search: the first maximum of the probability of finding the target state (\fref{fig:figura4}.a) is reached in correspondence of the first local minimum of entropy (\fref{fig:figura4}.b): that the search has gone, before this instant, through a local \underline{maximum} of entropy is shown with particular evidence by the Bloch diagram of \fref{fig:figura4}.c.
\section{The role of initial conditions}
An initial condition of the form
\begin{equation}
	\ket{R(1);\psi_0} = \ket{R(1)} \otimes \sum_{x=1}^{\epsilon} \psi_0(x)\ket{C(x)}
\end{equation}
with $\psi_0$ having support in a bounded region $\Lambda_{\epsilon} = \{1,2,\ldots,\epsilon\} \subseteq \{1,2,\ldots,s\}$ evolves, under \eref{eq:hamiltoniana2} as
\begin{equation}
	e^{-itH} \ket{R(1);\psi_0} = \sum_{x=1}^{\epsilon} \psi_0(t,x) \ket{R(x)}\otimes \ket{C(x)},
\end{equation}
where $\psi_0(t,x)$ solves, with the obvious boundary and initial conditions, the (discretized) free Schr\"odinger equation. The ensuing spreading of the wave packet  leads to an increasing trend (with the exception of the effects of reflection at time $t \approx s$ evidenced in figures \ref{fig:figura4}.b and \ref{fig:figura2}) of the von Neumann entropy $S(\rho_r(t))$ of the state
\begin{equation}
	\rho_r(t)=\sum_{x=1}^s \left| \psi_0(t,x)\right|^2 \ketbra{R(x)}{R(x)} 
\end{equation}
of the register. This is an undesirable feature because $S(\rho_r(t))$ gives a lower bound on the Shannon entropy of the distribution of any observable of the register, for short on the uncertainty in any reading of the output.\\
The models of the previous section where intended to show the above effect; in this section we devote some effort to the goal of decreasing it, by suitable choices of initial condition aimed at reducing the spreading of $Q$  in the state $\psi_0(t,x)$. It is sufficient, for \emph{this} purpose, to study only the cursor, evolving under the Hamiltonian
\begin{equation} \label{eq:freehamiltonian}
	H_0 = - \frac{\lambda}{2} \sum_{x=1}^{s-1} \tau_+(x+1) \tau_-(x)+\tau_+(x) \tau_-(x+1).
\end{equation}
The point is to devise an initial condition $\psi_0$  which uses  whatever additional finite amount $\Lambda_{\epsilon}= \{1,2,\ldots,\epsilon\}$ of space resources is available as a \emph{launch pad} for the cursor in an ``efficient'' way:  this means both a high value of the expectation of $V(\psi_0)$ and a small value of the variance of $V(\psi_0)$ (we want the spreading of $Q$ to increase at a low rate for a short time of computation). That both goals can be achieved is shown by examining the family of initial conditions, given by the eigenstates of a Hamiltonian of the form \eref{eq:freehamiltonian} restricted to qubits in $\Lambda_{\epsilon}$:
\begin{equation}
	\ket{c_k}= \sum_{x=1}^{\epsilon} \sqrt{ \frac{2}{\epsilon + 1} }\sin \left( \frac{k \pi x}{\epsilon+1}\right) \ket{C(x)},\; k=1,2,\ldots,\epsilon.
\end{equation}
The probability density of the speed $V_k \equiv V(c_k)$ corresponding to each of the above states is easily computed from \eref{eq:densitapsi0}:
\begin{eqnarray} \label{eq:densitavk}
	f_{V_k}(v)& = & I_{(0,1)}(v) \cdot  \\
	& \cdot & \frac{4 \left(3-2 v^2 + \cos \left( \frac{2 k \pi}{\epsilon + 1}\right) \right) \left( \sin \left( \frac{k \pi}{\epsilon + 1} \right) \right)^2 \left( \sin((\epsilon+1) \arcsin(v))\right)^2}{\pi \sqrt{1-v^2}(\epsilon + 1) \left( 2 v^2 + \cos \left( \frac{2 k \pi}{\epsilon + 1 }\right)-1\right)^2.  \nonumber}
\end{eqnarray}
\begin{figure}[htbp]
	\centering
		\includegraphics[width=300pt]{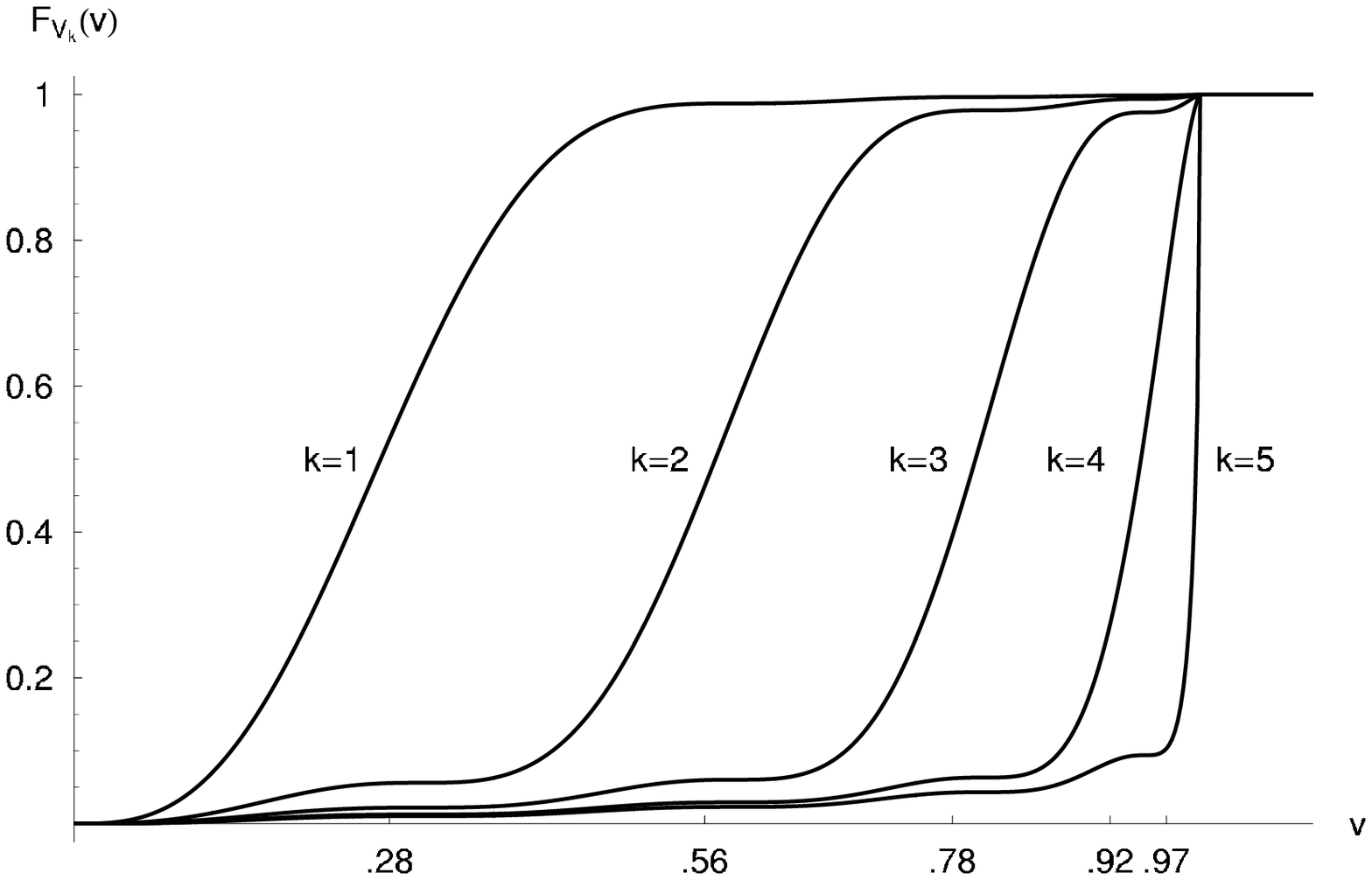}
	\caption{$n=5,\ \epsilon=2n-1=9$. The cumulative distribution functions  \mbox{$F_{V_k}(v)=Prob(V_k \leq v)$} corresponding to the densities \eref{eq:densitavk}, for $k$ going from $1$ to $n$. The ticks on the $v$ axes are $E(V_1)<E(V_2) < \ldots <E(V_5)$.}
	\label{fig:figura13}
\end{figure}
The behavior of $V_k$ is examplified by \fref{fig:figura13}. We are taking there, as we will always do in this section for the sake of notational convenience, $\epsilon$ to be odd
\begin{equation}
	\epsilon=2 n - 1.
\end{equation}
The examples of \fref{fig:figura13} clearly show the dispersive  nature of the medium \eref{eq:freehamiltonian}; they also show that increase of the mean value is accompanied by decrease of the variance (as shown by the increase in the steepness of the graph as $k$ goes from $1$ to $n$). An obvious choice for the initial state of the cursor emerges from the above example:
\begin{eqnarray} \label{eq:condiniflat}
	\ket{c_n} & = & \sum_{x=1}^{2n-1} \sqrt{\frac{1}{n}} \sin\left(\frac{\pi}{2} x \right) \ket{C(x)} =  \nonumber \\
	&=& \frac{\ket{C(1)}-\ket{C(3)}+\ket{C(5)}+ \ldots - (-1)^n \ket{C(2n-1)}}{\sqrt{n}} ,
\end{eqnarray}
conforming to the idea of packing the maximum number of wavelengths in the \emph{launch pad} $\Lambda_{\epsilon} = \{1,2,\ldots,\epsilon\}$, and having a, presumably easy to prepare, stationary state of the free  $XY$ chain localized in  $\Lambda_{\epsilon}$.\\
The random variable $V_n \equiv V(c_n)$  has probability density
\begin{equation}
	f_{V_n} = I_{(0,1)}(v) \frac{\left( \sin(2 n \arcsin(v))\right)^2}{ \pi n (1-v^2)^{3/2}}
\end{equation}
and, therefore, expectation value
\begin{eqnarray} \label{eq:attesavn}
	E(V_n) & = & \frac{4}{\pi}\sum_{h=1}^n \left(\frac{1}{ 4h-3}-\frac{1}{4h-1} \right) =    \nonumber \\
	& = & 1- \frac{4}{\pi}\sum_{h=n+1}^{+\infty} \left(\frac{1}{ 4h-3}-\frac{1}{4h-1} \right) \approx \nonumber \\
	& \approx & 1-\frac{1}{2 \pi n}
\end{eqnarray}
The second moment of $V_n$ is explicitly given by
\begin{equation}
	E(V_n^2) = 1-\frac{1}{4n}.
\end{equation}
The above considerations lead to the following asymptotic behavior, for large $n$, of the variance of $V_n$:
\begin{equation} \label{eq:varianzavn}
	var(V_n) = \frac{4 -\pi}{4 \pi n}.
\end{equation}
Equation \eref{eq:attesavn} is a quantitative assessment of the cost in terms of space resources of achieving the first requisite of \emph{efficiency}, namely high mean speed;  similarly, \eref{eq:varianzavn} gives the cost of decreasing the variance of $V_n$.\\
Incidentally, as the observable $Q$ has, in the state \ket{c_n}, expectation value
\begin{equation}
	E(Q_n) = \bra{c_n} Q \ket{c_n} = n
\end{equation}
and variance
\begin{equation}
	var(Q_n)=\frac{n^2-1}{3},
\end{equation}
equation \eref{eq:varianzavn} can be read as saying that, in the initial state \ket{c_n}, the position-velocity uncertainty product is given by
\begin{equation}
	var(Q_n)var(V_n) \approx \frac{n (4-\pi)}{12 \pi}
\end{equation}
Figures \ref{fig:figura5} and \ref{fig:figura7} show the relevance of the above \emph{asymptotic} considerations for the case of \emph{finite} $\epsilon$  and finite $s$ for $t<s$.
\begin{figure}[t]
	\centering
		\includegraphics{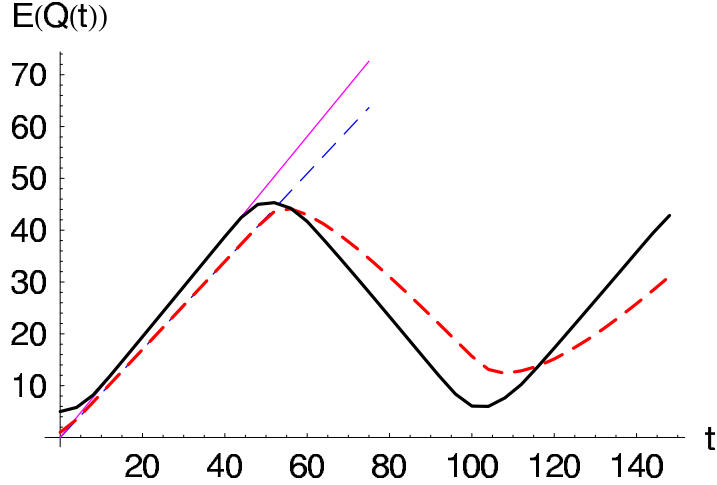}
	\caption{Solid lines: the expectation value of $Q$ in  a state \ket{M_n(t)} evolving from an initial condition having the cursor in  \ket{c_n}; the slope of the initial linear part of the graph is correctly predicted by \eref{eq:attesavn}. For comparison purposes the dashed lines show $\bra{M_1(t)}Q \ket{M_1(t)}$ as a function of time and the corresponding linear fit with slope given by \eref{eq:EM1}.}
	\label{fig:figura5}
\end{figure}
\begin{figure}[htbp]
	\centering
	\subfigure[]{\includegraphics[width=7cm]{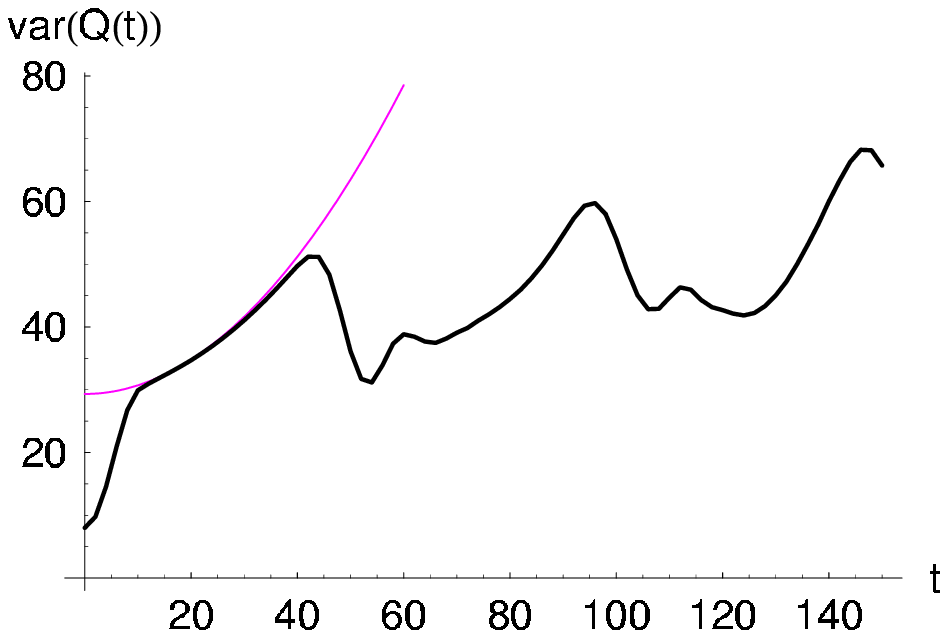}}
	\subfigure[]{\includegraphics[width=7cm]{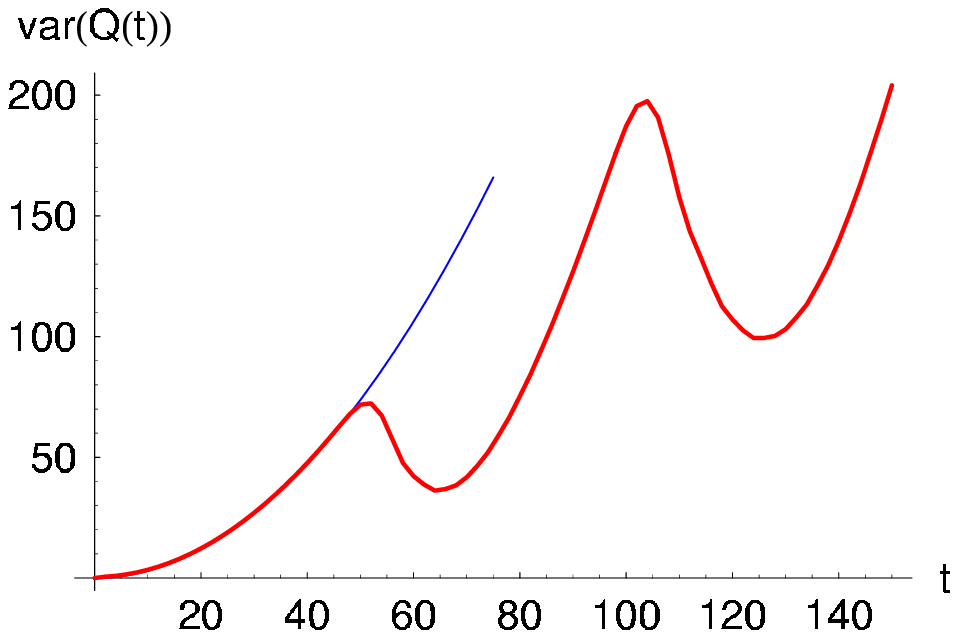}}
	\caption{$s=50,\ n=5$. (a) The variance of  $Q$ in a state \ket{M_n(t)} as a function of $t$, compared with its best fit of the form $const.+ t^2(4-\pi)/(4 \pi n)$, in the time interval $(\epsilon, s-\epsilon)$ in which boundary effects can be neglected. (b) The variance of $Q$ in the state \ket{M_1(t)}, compared with its approximation $t^2(3/4 -(8/(3 \pi))^2)$, suggested by \eref{eq:varM1}.}
\label{fig:figura7}
\end{figure}
The effect of the initial condition is most evident  if we compare the evolution of the state of the register from the initial state $\ket{M_1} = \ket{R(1)} \otimes \ket{C(1)}$ with the evolution starting from
\begin{equation}
	\ket{M_n} = \ket{R(1)} \otimes \ket{c_n}.
\end{equation}
\begin{figure}[ht]
\subfigure[]{\includegraphics[width=79mm]{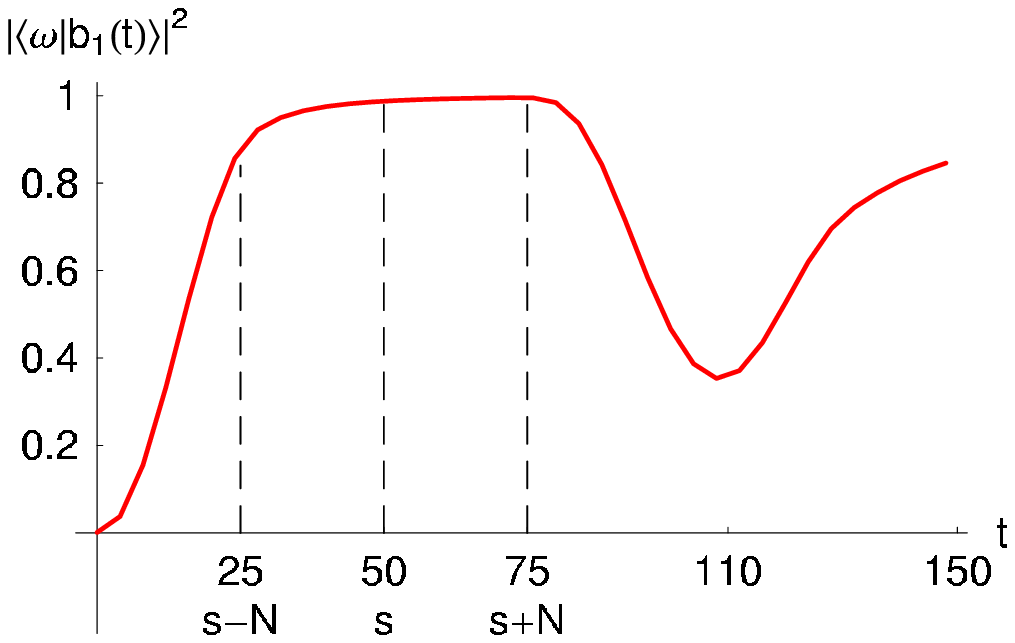}}
\subfigure[]{\includegraphics[width=79mm]{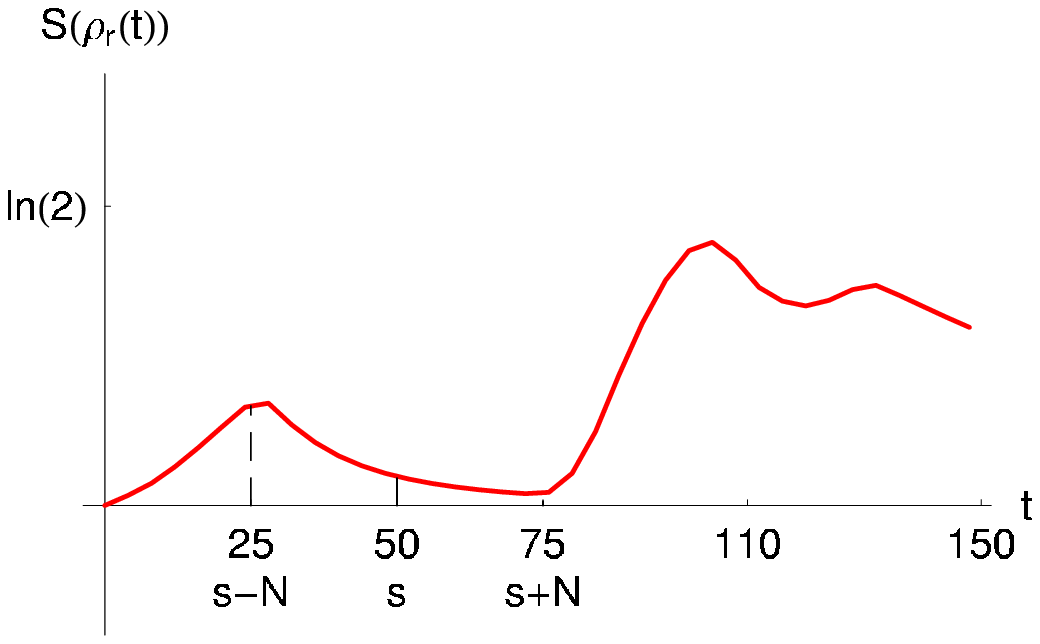}} \\
\centering
\subfigure[]{\includegraphics[width=59mm]{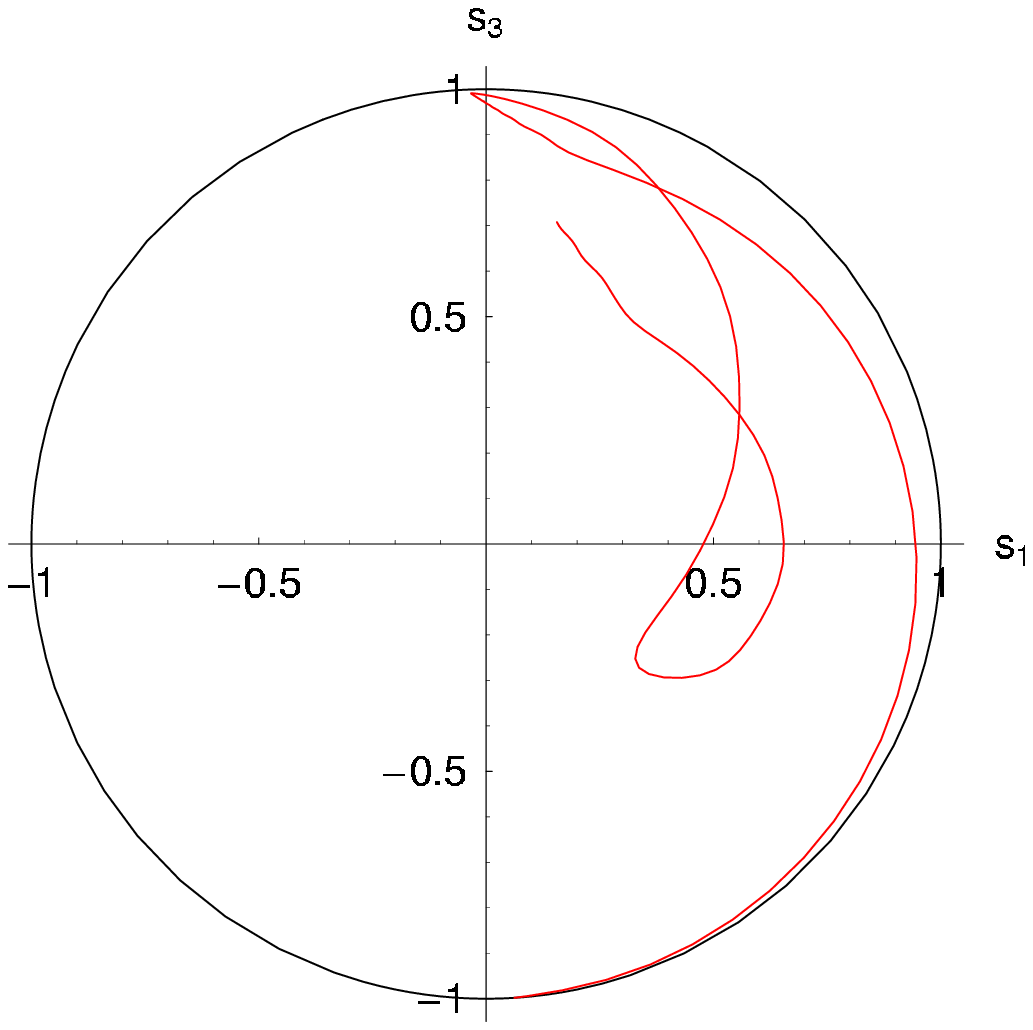}} 
\caption{$s=50,\ 0 \leq t \leq 3s,\ \mu=10,\ N= \left\lfloor \frac{\pi}{4} 2^{\mu/2} \right\rfloor$ (the Grover-optimal number of active steps); $U_1=U_2=\ldots=U_N=\exp(-i \alpha \sigma_2/2)$, with $\alpha$ and $\theta$ chosen as in \eref{eq:theta} and \eref{eq:alpha}; $U_x=I_r$ for $x > N$; initial state $\ket{M_1}= \ket{R(1)} \otimes \ket{C(1)}$.}
\label{fig:figura8}
\end{figure}
\begin{figure}[ht]
\subfigure[]{\includegraphics[width=79mm]{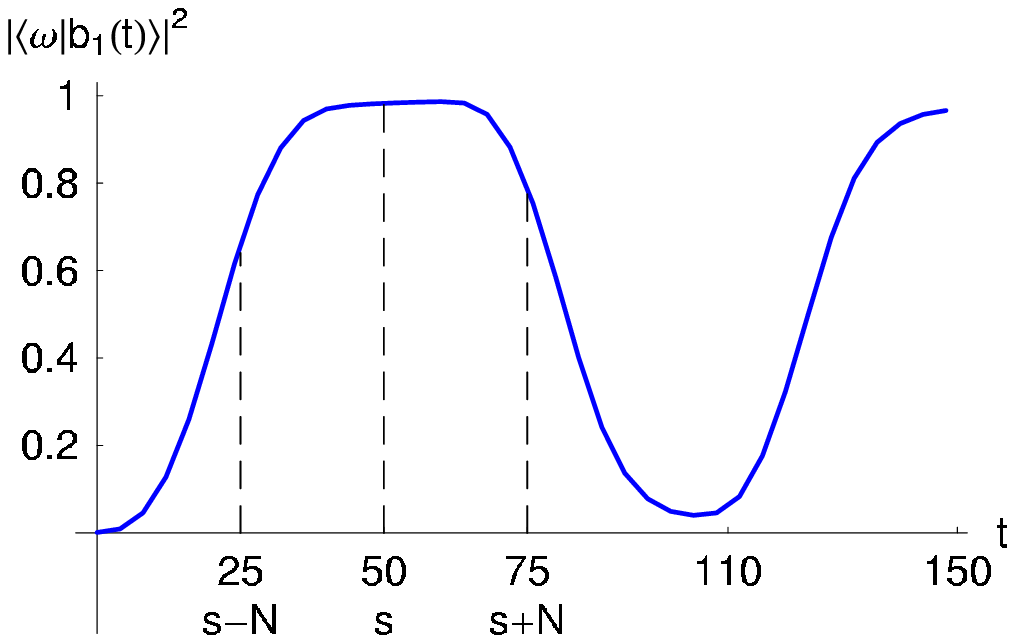}}
\subfigure[]{\includegraphics[width=79mm]{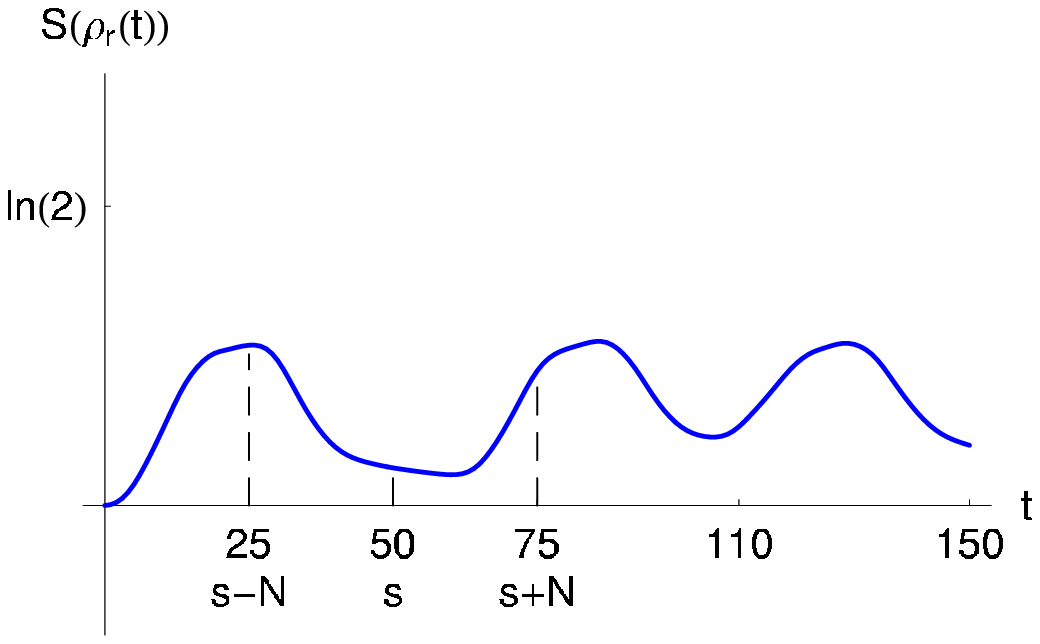}} \\
\centering
\subfigure[]{\includegraphics[width=59mm]{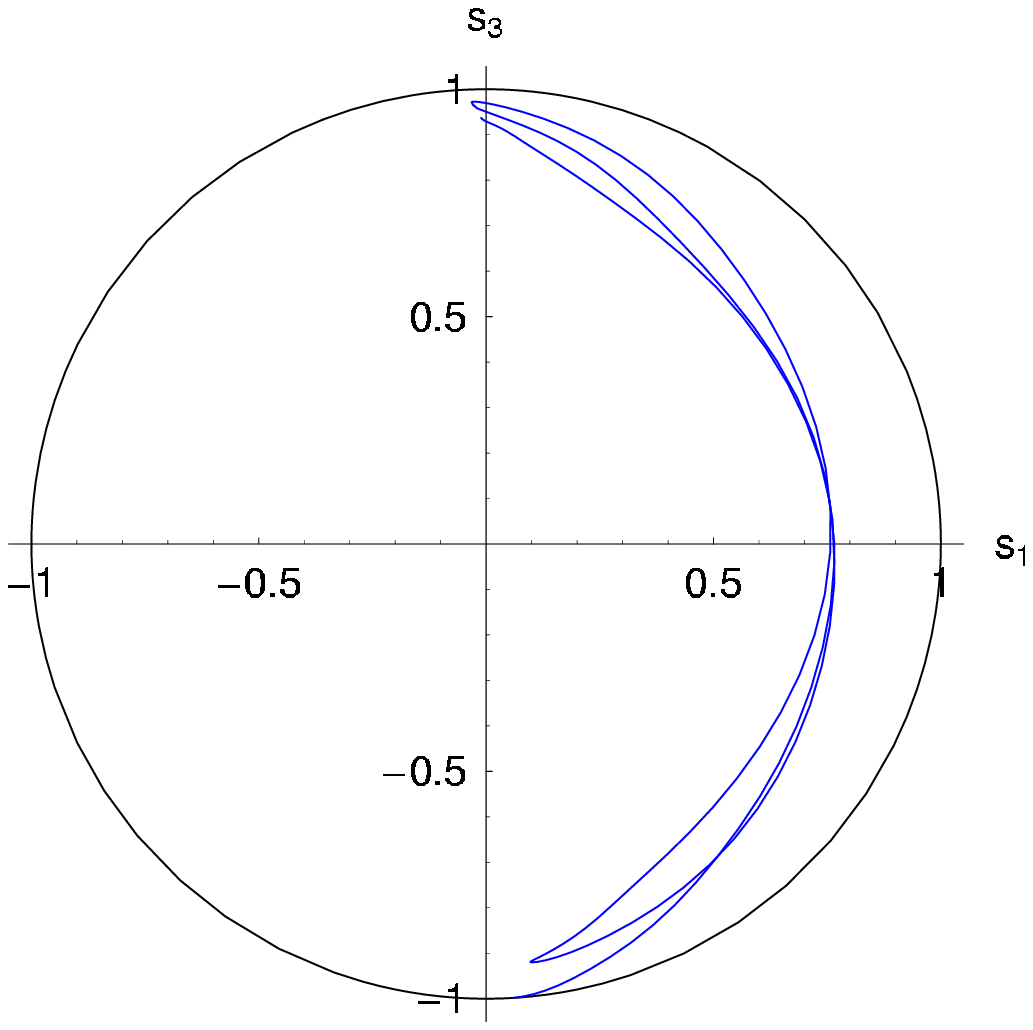}} 
\caption{$s=50,\ 0 \leq t \leq 3s,\ \mu=10,\ N=25,\ \alpha$ and $\theta$ as in \fref{fig:figura8}; $n=5,\ \epsilon=2n-1$; $U_{\epsilon}=U_{\epsilon+1}=\ldots = U_{\epsilon+N-1}= \exp(-i\alpha \sigma_2/2)$; $U_x=I_r$, for $1\leq x < \epsilon$ or $x \geq \epsilon + N$; initial state $\ket{M_n}= \ket{R(1)}\otimes \ket{c_n}$.}
\label{fig:figura9}
\end{figure}
This is done in figures \ref{fig:figura8} and \ref{fig:figura9} in the same \emph{probability-entropy-Bloch } format as in \fref{fig:figura4}. We examine there two different ways of using an additional amount $s-N$ of space, of size comparable with the minimum amount $N$ required by the algorithm. \Fref{fig:figura8} summarizes the experience developed in \cite{apolloni02} on the effect of using all this additional space as a telomeric chain or ``landing strip'': as long as the cursor stays in this region the register remains acted upon by the optimal number of primitives. \Fref{fig:figura9} shows the improvement obtained by investing part of the additional space as a ``launch pad'' on which to prepare a state in which the spreading of the cursor  increases (see \fref{fig:figura7}) at a lower rate than when starting from position 1.\\
Comparison of figures \ref{fig:figura8}.c and \ref{fig:figura9}.c, in particular the improvement of the behavior after  reflections at site $s$, shows that the idealized scenario of reversible computation (the cursor, ``going back and forth'', ``does and undoes'' the reversible computation) is within reach, with, as \eref{eq:varianzavn} shows, a polynomial cost in space.
\begin{figure}[ht]
\subfigure[]{\includegraphics[width=79mm]{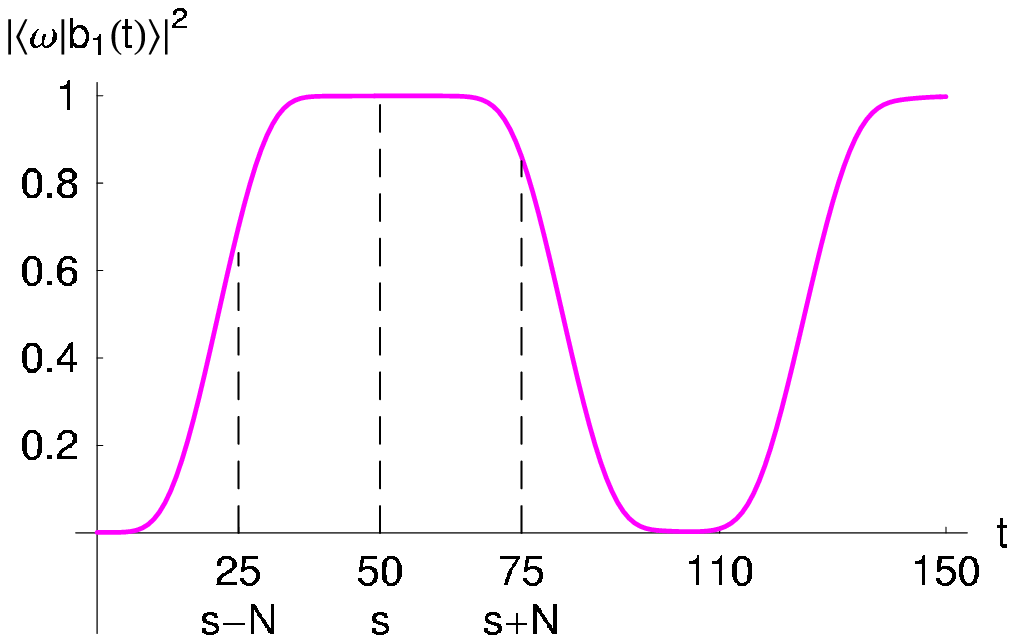}}
\subfigure[]{\includegraphics[width=79mm]{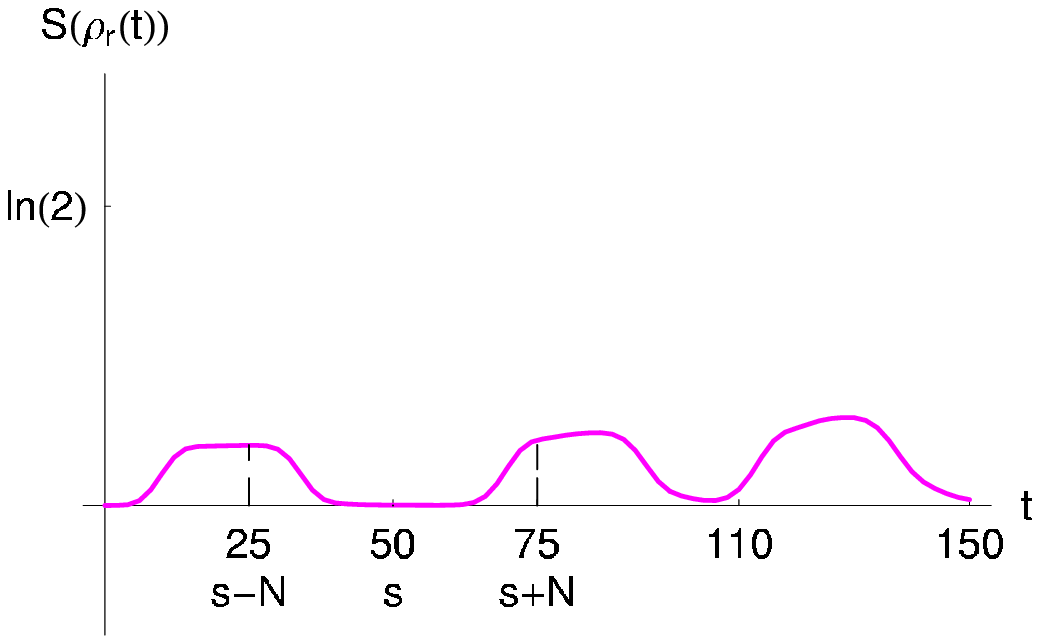}} \\
\centering
\subfigure[]{\includegraphics[width=59mm]{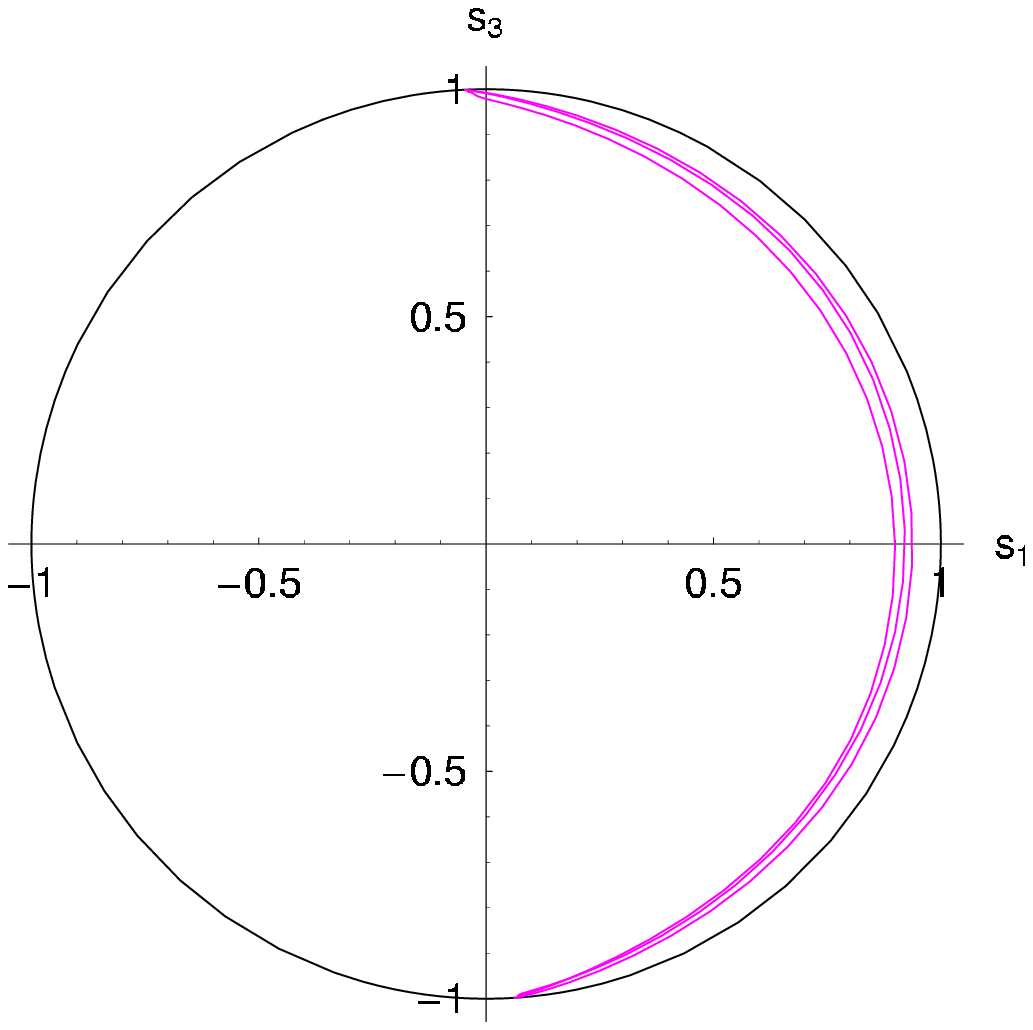}} 
\caption{Same parameters as in \fref{fig:figura9}; initial state  $\ket{R(1)}\otimes \ket{\gamma_n}$, as in \eref{eq:gamman}.}
\label{fig:figura10}
\end{figure}
We note, in \fref{fig:figura10}, that we can do much better than in \fref{fig:figura9}, with the same expenditure of space resources, in approximating the reversible scenario if, instead of the initial state \eref{eq:condiniflat}, we set the cursor in the initial state
\begin{equation} \label{eq:gamman}
	\ket{\gamma_n}=\sqrt{\frac{2}{3n}} \sum_{x=1}^{2n-1} \left( 1+\cos \left( \frac{\pi}{2n} x\right)\right) \sin\left( \frac{\pi}{2} x\right) \ket{C(x)}.
\end{equation}
The state $\ket{\gamma_n}$ emerges quite naturally as a three-mode approximation (a linear combination of \ket{c_n}  and  \ket{c_{n\pm1}}) of the initial condition that maximizes the mean speed of computation for fixed length $\epsilon$ of the \emph{launch pad}.

\section{Number of particles}
In the previous section we have provided examples of the benefit of spreading the initial wave function of the cursor ($N_3=1$) on an initial \emph{launch pad} instead of, as it would be classically ``obvious'', having it strictly localized at site $1$. Equality \eref{eq:varianzavn} is, in this context, a quantitative assessment of the cost, in term of space resources, of implementing Feynman's ballistic mode of computation.\\
In  this section we  abandon, in the same spirit, the classical prejudice of having a single clocking excitation, and present a preliminary analysis of the idea of starting the cursor in an initial state with  $N_3>1$. The idea is to follow the motion of a swarm of several clocking agents (cursor spins in the ``up'' state) acting on the register. Stated otherwise, with  reference for simplicity to the case  $N_3=2$, we allow the \emph{clock} to perform a quantum  walk on the graph having the vertices $(x_1,x_2)$, with  $1 \leq x_1 < x_2 \leq s$, with edges between nearest neighbors \cite{osborne04}.\\
We recall, mainly in order to establish our  notation, a few elementary  facts \cite{lieb61} about the $XY$ Hamiltonian \eref{eq:freehamiltonian}.\\
The eigenstates of $H_0$ in the subspace $N_3=n$ are labeled by subsets of size $n$ of $\Lambda_s=\{1,2,,\ldots,s\}$; if $\mathsf{K}=\{\ves{k}{n}\}$ is such a subset (where we will always assume $1 \leq k_1 < k_2 < \ldots <k_n \leq s$), an eigenstate of $H_0$ belonging to the eigenvalue
\begin{equation} 
	E_{\mathsf{K}}=\sum_{j=1}^n e_{k_j}
\end{equation}
is given by
\begin{equation}
	\ket{E_\mathsf{K}}= \sum_{M \subseteq \Lambda_s; |M|=n} V(\mathsf{K},M) \ket{M}.
\end{equation}
For  $M=\{\ves{x}{n}\}$, with $1 \leq x_1 < x_2 < \ldots <x_n \leq s$, we have indicated above by \ket{M} the simultaneous eigenstate of $\ve{\tau_3}{s}$  in which only the spins in $M$ are ``up'', and we have set:
\begin{equation}
	V(\mathsf{K},M)= \det\left(\left\|v_{k_i}(x_j) \right\|_{i,j=1,\ldots,n}\right)
%\begin{array}{c}
%	i=1,\ldots,n\\
%	j=1,\ldots,n
%\end{array}
\end{equation}
where the functions $v_k$ have been defined in \eref{eq:autofunzioni}.\\
We set
\begin{equation}
	Q_i \ket{\{\ves{x}{n}\}} = x_i \ket{\ves{x}{n}}.
\end{equation}
It is easy to study, by the techniques of \sref{sec:speed}, the asymptotic (as $s \to +\infty$ and $t \to +\infty$) joint distributions of the observables $Q_i$, and therefore to give quantitative estimates of the correlation between the speeds of different \emph{particles} and its dependence on the initial condition. To quote just one example, in the subspace $N_3=2$ and in the state \ket{\{1,2\}}  the velocities $(V_1,V_2)$ of the two ``up'' spins  (the limits in law of $Q_1/t$ and $Q_2/t$, respectively) have joint probability density
\begin{equation} \label{eq:jointdistr}
	f_{V_1,V_2}(v_1,v_2) = I_{(0,v_2)}(v_1) I_{(0,v_1)}(v_2)\frac{64 v_1^2 v_2^2 (2-v_1^2-v_2^2)}{\pi^2 \sqrt{(1-v_1^2)(1-v_2^2)}} 
\end{equation}
It is immediate from \eref{eq:jointdistr} to compute the conditional expectation  $E(V_1 | V_2)$ of the velocity of the leftmost \emph{particle} given the one of the rightmost \emph{particle}; it turns out to be:
\begin{equation}
	E(V_1|V_2)= \frac{3 V_2}{4}+ O(V_2^5). 
\end{equation}
In this section we advance the following idea: if the issue of the computation is the application, for a given number $g$ of times, of a given primitive $G$ to the register, initialize the cursor in the $N_3=g$ subspace, in the state, say, \ket{\{1,2,\ldots,g\}}; let then the system evolve according to the Hamiltonian:
\begin{equation} \label{eq:hamiltoniana5}
	H = -\frac{\lambda}{2} \sum_{x=1}^{s-1} U_x \otimes \tau_+(x+1) \tau_-(x)+  U_x^{-1} \otimes \tau_+(x) \tau_-(x+1)
\end{equation}
where
\begin{equation} \label{eq:ux}
	U_x = G^{\delta_{x_0,x}},\mbox{ for a fixed $x_0 \geq g$},\; G^0 = I_r.
\end{equation}
\begin{figure}[ht]
\subfigure[]{\includegraphics[width=79mm]{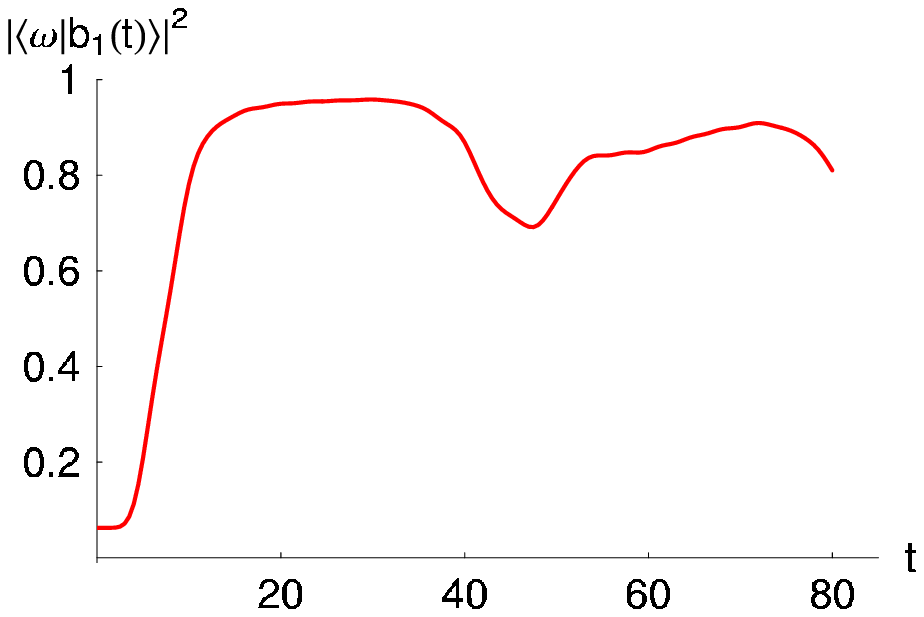}}
\subfigure[]{\includegraphics[width=79mm]{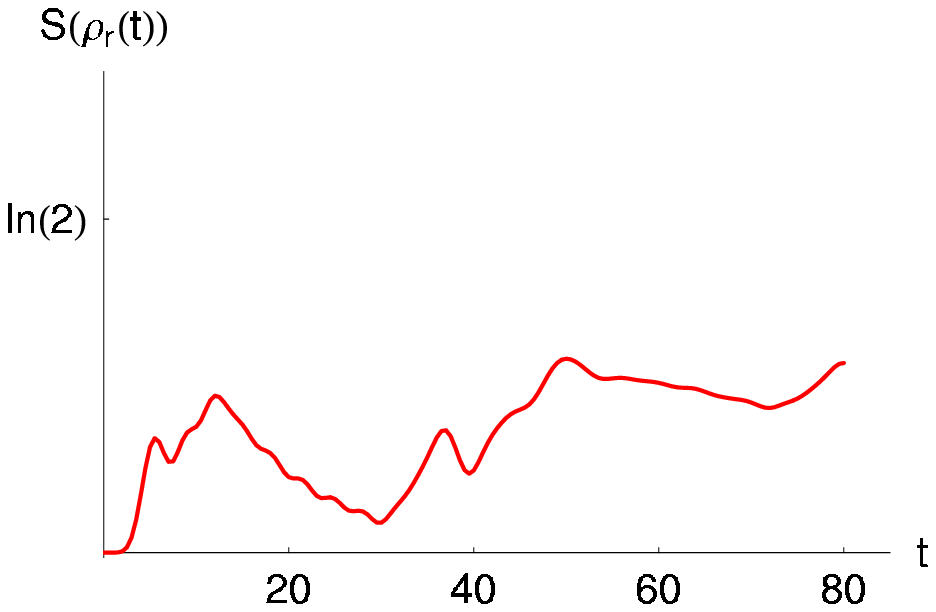}} \\
\centering
\subfigure[]{\includegraphics[width=59mm]{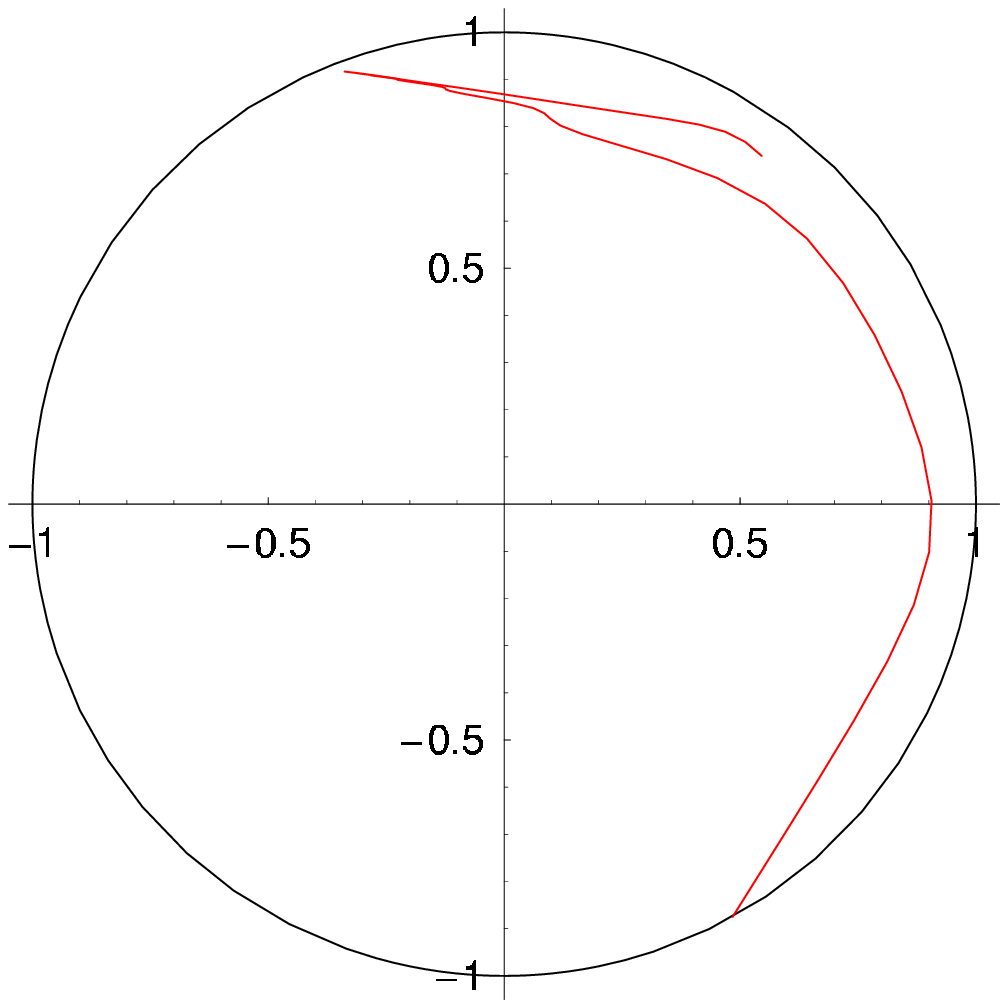}} 
\caption{$\mu=4,\ g=3,\ x_0=6,\ s=20,\ 0 \leq t \leq 4s$; $U_{x_0}=G=\exp(-i \alpha \sigma_2 / 2)$ with $\alpha$ and $\theta$ given by \eref{eq:alpha}  ad \eref{eq:theta}, $U_x=I_r$ for $x \neq x_0$; initial condition $\ket{R(1)} \otimes \ket{\{1,2,3\}}$ with \ket{R(1)} given by \eref{eq:condinigrover}.}
\label{fig:figura11}
\end{figure}
An implementation of this approach is shown by the probability-entropy-Bloch diagram of \fref{fig:figura11}.
Simple expressions for the quantities shown in \fref{fig:figura11} can be obtained by the explicit form of the eigenvectors of the Hamiltonian described by \eref{eq:hamiltoniana5} and \eref{eq:ux} in every eigenspace of $N_3$. For instance in the subspace $N_3=3$ a complete set of eigenstates is given, for $\zeta=\pm 1$ and $1 \leq j < k \leq s$, by:
\begin{eqnarray} \label{eq:evolutomulti}
	\ket{\zeta;E_{\{j,h,k\}}} & = & \sum_{1 \leq x_1 <x_2 < x_3 \leq s} V(\{j,h,k\},\{x_1,x_2,x_3\})\cdot \nonumber \\
	& \cdot &  G^{\vartheta(x_1-x_0)+\vartheta(x_2-x_0)+\vartheta(x_3-x_0)} \ket{\sigma_3=\zeta} \otimes \ket{\{x_1,x_2,x_3\}}
\end{eqnarray}
where $\vartheta$ is the unit step function defined by:
\begin{equation} \label{eq:thetax}
	\vartheta(x)= \left \{ 
	\begin{array}{c r}
	1, & \mbox{if $x > 0$}\\
	0, & \mbox{if $x \leq 0$} 
	\end{array}\right.
\end{equation}
\begin{figure}[htbp]
	\centering
	\subfigure[]{\includegraphics[width=7cm]{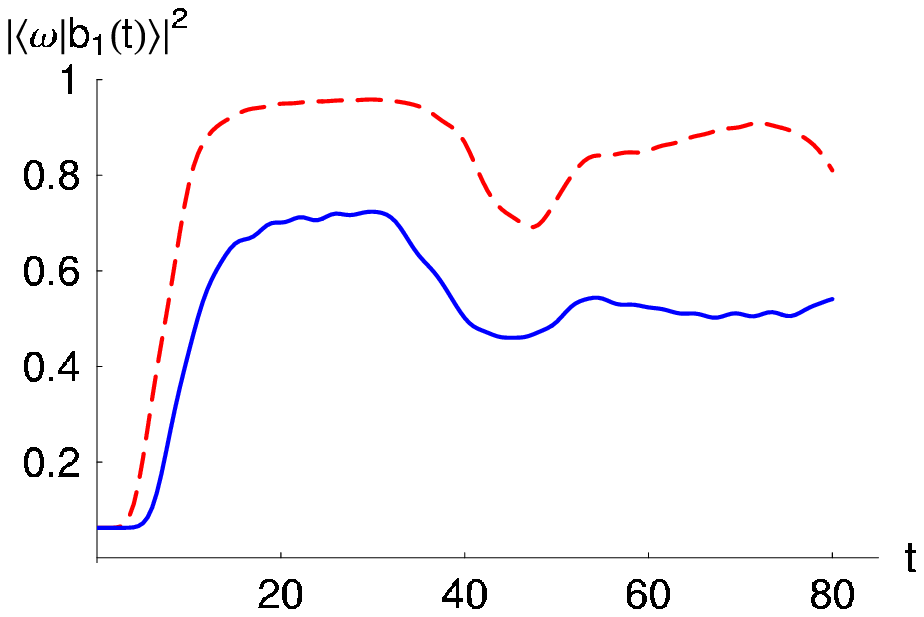}}
	\subfigure[]{\includegraphics[width=7cm]{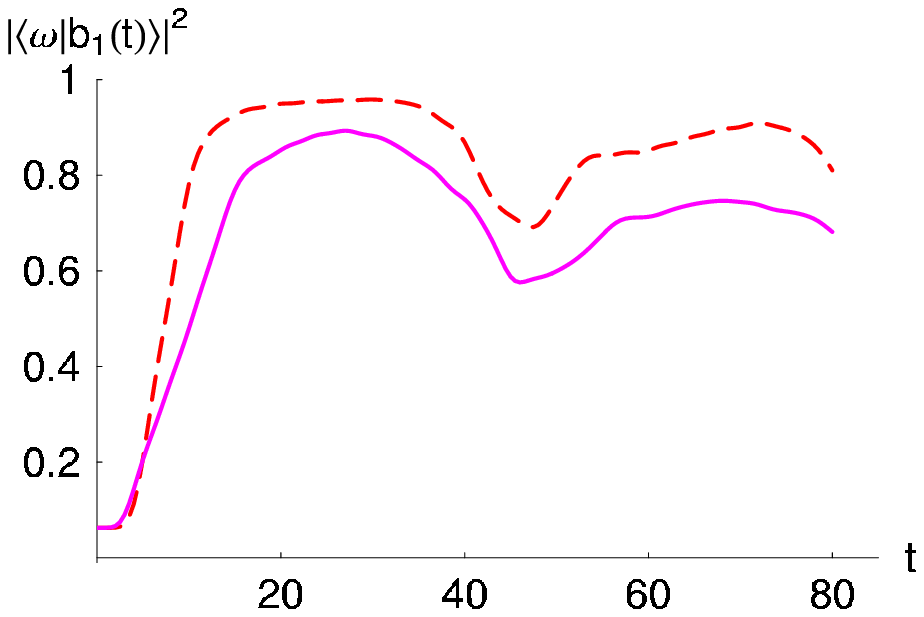}}
	\caption{$\mu=4,\ s=20,\ 0 \leq t \leq 4s$; (a) solid line: $U_a=A,\ U_b=B,\ a=6,\ b=8$, same initial condition as in \fref{fig:figura11}; (b) solid line: the $N_3=3$ state has been prepared by setting the initial chain $\{1,\ldots,6\}$ in its ground state. For comparison purpose \fref{fig:figura11}.a is reproduced in both frames as a dashed line.}
\label{fig:figura12}
\end{figure}
The spectral structure \eref{eq:evolutomulti} is peculiar of the extremely simple situation \eref{eq:ux} (just one active link) considered there. As soon as we have more than one active link, say the primitive $A$ acting on link $(a,a+1)$ and the primitive $B$ acting on link $(b,b+1)$, with $b>a+1$, a new phenomenon (that for simplicity we discuss in the $N_3=2$ case) takes place: the energy eigenstates have not anymore the form of a linear combinations of tensors products of the form  $M(x_1,x_2)\ket{\sigma_3=\zeta} \otimes \ket{\{x_1,x_2\}}$, with $M(x_1,x_2)$ a \underline{monomial} in $A$ and $B$; related to this, the coordinates $x_1,\ x_2$ lose, strictly speaking, the meaning of relational time \cite{gambini04b}: given that at a given value of $t$, $Q_1=x_1$ and $Q_2=x_2$ we can only claim that the state of the register has been acted upon by a \underline{polynomial} in $A$ and $B$.\\
This phenomenon is easily understood in terms of the Dyson expansion of the propagator: the probability amplitude for the two excitations being in $x_1,\ x_2$ (both larger than $b$), given that at time $0$ they were in $y_1,\ y_2$ (both $\leq a$), receives contributions not only from Feynman paths along which the rightmost excitation goes past $a$ and $b$ and \underline{then} the leftmost excitation goes past  $a$ and $b$ (along such a computational path the state of the register is modified by $BABA$), but also, among others, from paths along which both excitations go past $a$ before both going past $b$ (along such a computational path the state of the register is modified by  $BBAA$).\\
Waiting for an algorithm that might benefit from the above possibility of simultaneously exploring different computational paths (concurrency ?), we explore, in \fref{fig:figura12}, the idea (or classical prejudice?) that this nuisance  can be in part avoided by using suitable initial conditions. The idea, suggested by \eref{eq:densitavk}, is of course to prepare on $\Lambda_a=\{1,2,\ldots,a\}$ an initial $N_3=g$ state such that the excitations travel as spatially well localized wave packets of so different speeds that it is at any time unlikely that they simultaneously hit the region $(a+1,b)$.\\
We conclude with a remark about our insistence, throughout the paper, in gathering experience about the  behavior of the evolution of a state of an initial subchain \mbox{$\Lambda_{\epsilon}= \{1,2,\ldots,\epsilon\}$}.\\
We observe that an initial state  (not necessarily in the $N_3=1$ subspace) in $\Lambda_{\epsilon}$ of the form
\begin{equation}
	\ket{in}= \frac{1}{2^{\epsilon}} \sum_{M \subseteq \Lambda_{\epsilon}} \left( \sum_{z \in \{-1,1\}^{\epsilon}} f(z) \prod_{j \in M} z_j\right) \ket{\tau_3(x)=(-1)^{I_M(x)}},
\end{equation}
%\mbox{ iff } x\in M,\
where $I_M$ is the indicator function of the set $M$ and $x=1,\ldots,\epsilon$, can be prepared as a post-kickback state (with respect to an ancilla qubit) after the reversible evaluation of a function $f:\{-1,1\}^{\epsilon} \to \{-1,1\}$. We conjecture that subsequent evolution of \ket{in} under the Hamiltonian \eref{eq:freehamiltonian} on $\Lambda_{s}= \{1,2,\ldots,s\}$, with $s>>\epsilon$, might help in setting tests of hypotheses about the Fourier coefficients
\begin{equation}
	c_M=\frac{1}{2^{\epsilon}} \sum_{z \in \{-1,1\}^{\epsilon}}f(z) \prod_{j \in M} z_j
\end{equation}
of the function $f$ via time-of-flight techniques. There is at least one non trivial case in which the above conjecture works: having prepared all spins in  $\{\epsilon+1,\ldots,s\} $ in the ``up'' state, the Deutsch-Josza alternative \cite{deutsch92} ``constant ($c_{\emptyset}=1$) vs.   balanced ($c_{\emptyset}=0$)'' becomes equivalent to the alternative ``stationary vs. non stationary'' under the Hamiltonian  \eref{eq:freehamiltonian},  about the state of the overall system.
\section{Conclusions and outlook}
The pure $XY$ Hamiltonian $H_0$  given in \eref{eq:freehamiltonian} describes, in the Luther-L\"uscher-Susskind formalism \cite{luther76,luscher76,susskind77}, a massless Dirac quantum field on a $1$-dimensional lattice. The full Hamiltonian \eref{eq:hamiltoniana3} is suggestive of the minimal coupling of this Fermi field, implementing the \emph{clock}, with additional quantum fields implementing the \emph{register}. This work intends to contribute to the line of research, that seems to be emerging these days \cite{childs04, strauch05, verstraete05}, devoted to making this connection between quantum computing and relativistic quantum field theory explicit. It is an easy guess that this quantum field theoretical intuition was well present in the original work \cite{feyn86}. Particularly penetrating is, in this respect, Peres' remark that in Feynman's model \emph{calculations run forward and backward in time just as particles and antiparticles in Feynman's classical work on relativistic quantum field theory} (\cite{peres85}, p. 3269). As a further remark, we observe that the three-body interactions needed by Feynman's model are hard to conceive out of a field theoretical context.\\
It is because of this field theoretical perspective that we have tried to avoid any ``engineering'' (space dependence) of the coupling constant $\lambda$ in \eref{eq:freehamiltonian}, well aware of the fact that, in the $Dirac \to XY$  correspondence, $\lambda$ is related to the spacing adopted in the lattice approximation. In such a context it would be very hard to understand (without a projection mechanism \cite{christandl04}, which seems to have an exponential cost) the implementation of a space dependence such as
\begin{equation} \label{eq:lambdax}
	\lambda(x) = const.\, \sqrt{x(x-s)}
\end{equation}
that leads in \cite{peres85} and \cite{christandl04} to the existence of sharply distinguished instants in which the position of the cursor is certain. Nor would it be easy to understand \eref{eq:lambdax} in a solid state implementation \cite{burkard04}, where $\lambda$ is related to the effective mass of the clocking excitation.\\
In this paper we have focused our attention on the \emph{clocking field} $\tau(x)$, singled out as the one which, under suitable boundary conditions and for initial conditions localized close to the boundary, exhibits particle-like excitations performing, for long enough intervals of $t$, a quantum walk in a distinguished direction.\\[5pt]
Spatial homogeneity of the chain leads to the existence of the limit in law \mbox{$V=\lim_{t \to  +\infty} Q(t)/t$} for the position of such an excitation on a semi-infinite ($s \to +\infty$) box. In the $N_3=1$ subspace, because of Peres' conservation law \cite{peres85}, the observable $Q$ acquires the meaning of \emph{relational time} (\underline{given} the observed value of $Q$, the state of the register is known with certainty) and, therefore, the random variable $V$ acquires the meaning of number of computational steps per unit $t$. The fact that the variance of $V$ is strictly positive has the effect that in terms of the \emph{parameter time} $t$ (as opposed to \emph{relational time} $Q$) the evolution of the register appears to be dissipative: we have, for a simple model, written the corresponding Lindblad evolution and studied the ensuing build-up of entropy.\\
On our simple instance of quantum search we have shown that, in the ``low level'', physical approach that we pursue (in which time runs, for the cursor, because it is coupled with an additional quantum field) the build-up of entropy imposes an upper bound on the probability of finding the target state which is more severe than the one predicted by the ``high level'', algorithmic approach (in which the successive primitives are applied by an external macroscopic agent).\\
In the attempt of decreasing the deficit in the probability of success in a quantum search, due to the decohering effect of the coupling with the clocking field, we have provided examples of the benefit of spreading the initial wave function of the cursor on an initial \emph{launch pad} instead of, as a classical prejudice would suggest, having it strictly localized at one site.\\
We have, similarly, abandoned the classical prejudice of having a single clocking excitation, providing a preliminary analysis of the idea of starting the cursor in an  initial state with $N_3 > 1$. We have shown, in this context, an efficient way of iterating the application of a single primitive to the register and experienced the possibility, by a suitable choice of the initial conditions, of steering the quantum walk of the excitations in such a way as to reduce the conflicts about the order of application of non commuting primitives.\\[5pt]
The case $N_3>1$ deserves, we think, further research, both from the algorithmic and the physical point of view.\\ 
From the algorithmic point of view we plan to examine other instances (beyond the one cursorily examined at the end of section 6) in which time-of-flight spectroscopy (based on the Fourier transform vs. speed relationship recalled in section 3) of the post-kickback state can answer Yes/No questions about the algorithm.\\
From the physical point of view, the ``obvious'' choice of the ``all down'' reference state made throughout the paper is far from being optimal from the point of view of studying the thermodynamic cost of resetting the register. The best reference state for the study of this ultimate cost of reversible computation would of course be the ground state and, for Hamiltonians of the form \eref{eq:freehamiltonian}, with $s$ even, it is an $N_3=s/2$ state. This will require, we think, the formulation of an appropriate Bethe Anzatz for the Hamiltonian \eref{eq:hamiltoniana3}.
%(for example, the ground state the Hamiltonian \eref{eq:freehamiltonian} with even $s$, is an $N_3=s/2$ state). This will require, we think, the formulation of an appropriate Bethe Anzatz for the \mbox{Hamiltonian \eref{eq:hamiltoniana3}}.
%%%%%%%%%%%%%%%%%%%%%%%%%%%%%%%%%%%%%%
%%%%%%%%%%%%%%%%%%%%%%%%%%%%%%%%%%%%%%
\section*{Acknowledgement}
We wish to thank professors Alberto Bertoni and Massimiliano Goldwurm for stimulating discussions throughout the development of this work.
%\section*{References}
\bibliography{SpeedandEntropy}
\bibliographystyle{abbrv}
\end{document}